\documentclass[openacc]{rsproca_new}

\usepackage{overpic}
\usepackage{tikz}
\newcommand{\beq}{\begin{equation}}
\newcommand{\eeq}{\end{equation}}
\newcommand{\lb}{\left(}
\newcommand{\rb}{\right)}
\newcommand{\un}[1]{\,\mathrm{#1}}
\usepackage{url}

\usepackage{comment}
\usepackage{xr}
\usepackage{empheq}
\newcommand*\circled[1]{\tikz[baseline=(char.base)]{
            \node[shape=circle,draw,inner sep=2pt] (char) {#1};}}
\begin{document}

\title{Brachistochrone on a velodrome}

\author{
GP Benham$^{1}$, C Cohen$^{1}$, E Brunet$^{2}$ and C Clanet$^{1}$}

\address{$^{1}$LadHyX, UMR CNRS 7646, Ecole polytechnique, 91128 Palaiseau, France\\
$^{2}$F\'{e}d\'{e}ration Fran\c{c}aise de Cyclisme, 1 Rue Laurent Fignon, 78180 Montigny-le-Bretonneux, France}

\subject{Applied mathematics, Sports physics, Fluid dynamics, Biomechanics}


\corres{GP Benham\\
\email{graham.benham@ladhyx.polytechnique.fr}}

\begin{abstract}
The Brachistochrone problem, which describes the curve that carries a particle under gravity in a vertical plane from one height to another in the shortest time, is one of the most famous studies in classical physics. There is a similar problem in track cycling, where a cyclist aims to find the trajectory on the curved sloping surface of a velodrome that results in the minimum lap time. In this paper we extend the classical Brachistochrone problem to find the optimum cycling trajectory in a velodrome, treating the cyclist as an active particle. Starting with two canonical cases of cycling on a sloping plane and a cone, where analytical solutions are found, we then solve the problem numerically on the reconstructed surface of the velodrome in Montigny le Bretonneux, France. Finally, we discuss the parameters of the problem and the effects of fatigue.
\end{abstract}



\begin{fmtext}

\end{fmtext} \maketitle

\section{Introduction.}

In 1696 Johann Bernoulli posed a problem to the scientific community which, after a year and half, had only been solved by a handful of individuals, including Newton and Leibniz \cite{hand1998analytical, sussmann2002brachistochrone}. 
Along with Newton's minimal resistance problem, it was one of the first mathematical studies that pioneered the field of the variational calculus, and so had an immense influence thereafter \cite{goldstine2012history}.  
The Brachistochrone problem, whose etymology comes from the ancient Greek for \textit{shortest time} \cite{hand1998analytical}, describes a curve that carries a particle under gravity in a vertical plane from one height to another in minimal time. The solution is equivalent to the path traced out by a rolling circle, also known as a cycloid. Other variations of this problem have included the effects of friction \cite{ashby1975brachistochrone,vratanar1998analytical,hayen2005brachistochrone}, the motion of a disc on a hemisphere \cite{perantoni2013time},  the elastic Brachistochrone (elastochrone) \cite{aristoff2009elastochrone}, and even the quantum Brachistochrone problem \cite{carlini2006time}.

\begin{figure}
\centering
\begin{tikzpicture}[scale=0.5]
\node at (0,0) {\begin{overpic}[width=0.33\textwidth,trim={8cm 4cm 8cm 4cm},clip]{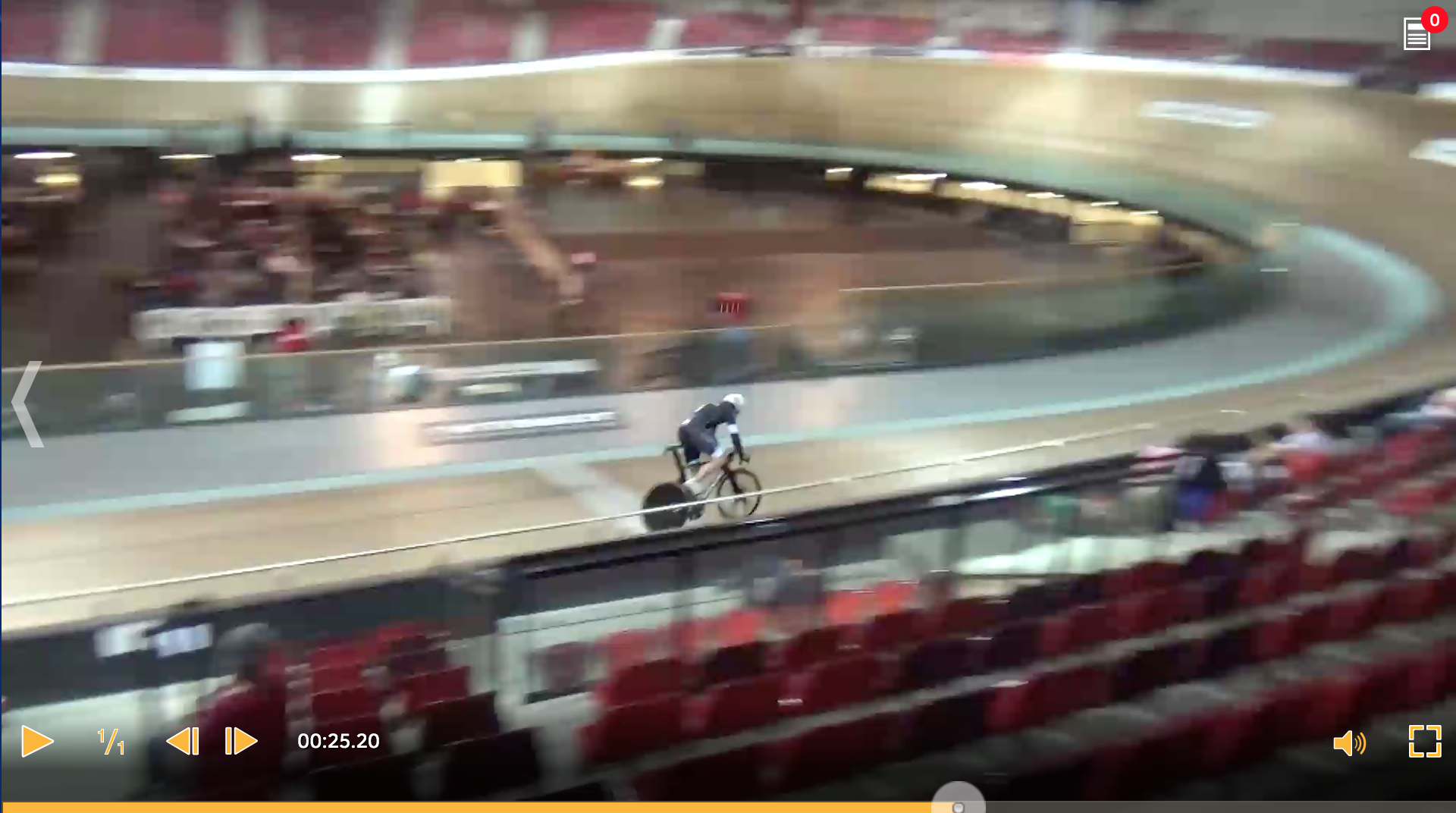}
\put (5,45) {\large \color{white} a)}
\end{overpic}};
\end{tikzpicture}
\hspace{0.5cm}
\begin{tikzpicture}[scale=0.5]
\node at (0,0) {\begin{overpic}[width=0.33\textwidth,trim={8cm 4cm 8cm 4cm},clip]{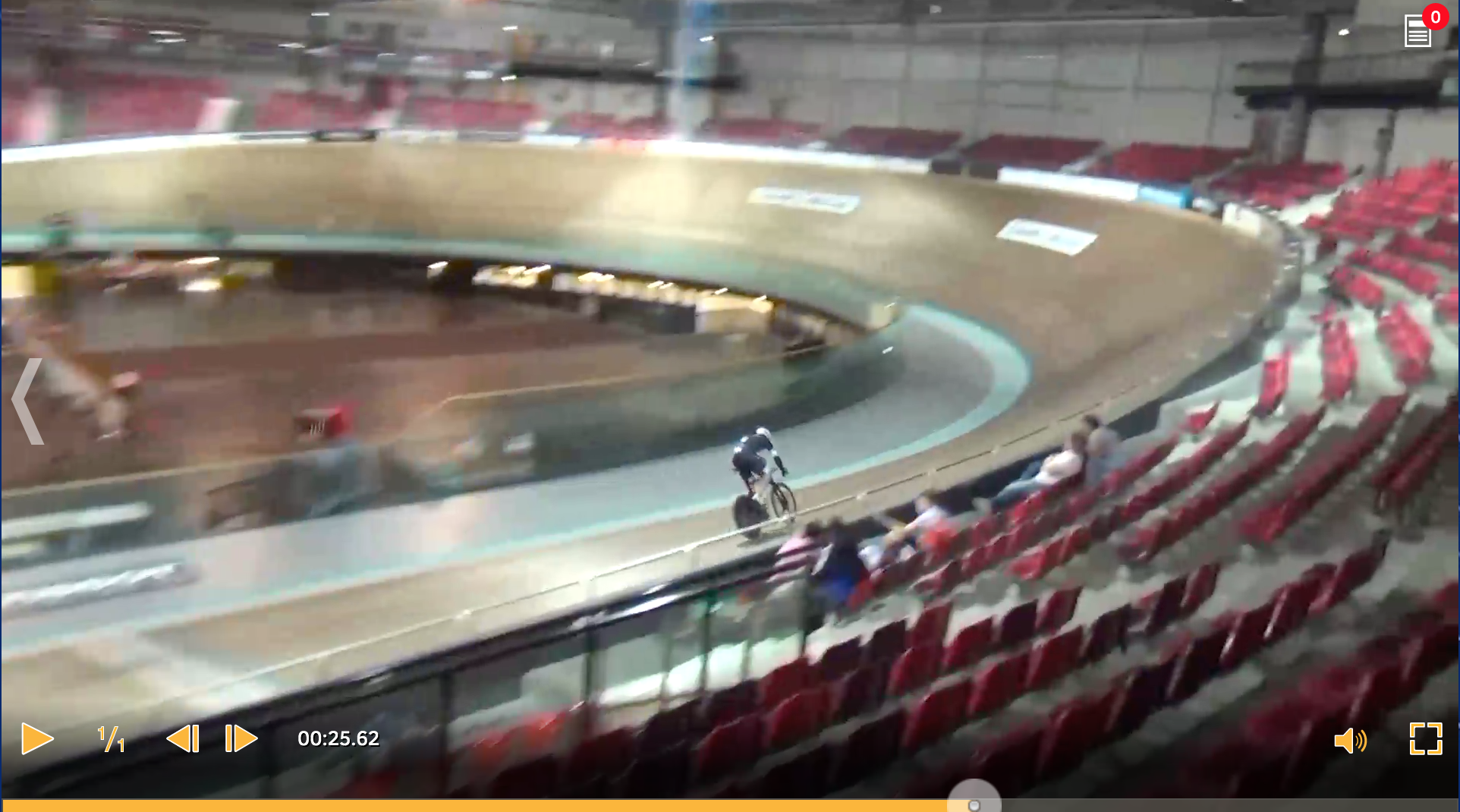}
\put (5,45) {\large \color{white} b)}
\end{overpic}};
\end{tikzpicture}\\
\vspace{0.5cm}
\begin{tikzpicture}[scale=0.5]
\node at (0,0) {\begin{overpic}[width=0.33\textwidth,trim={8cm 4cm 8cm 4cm},clip]{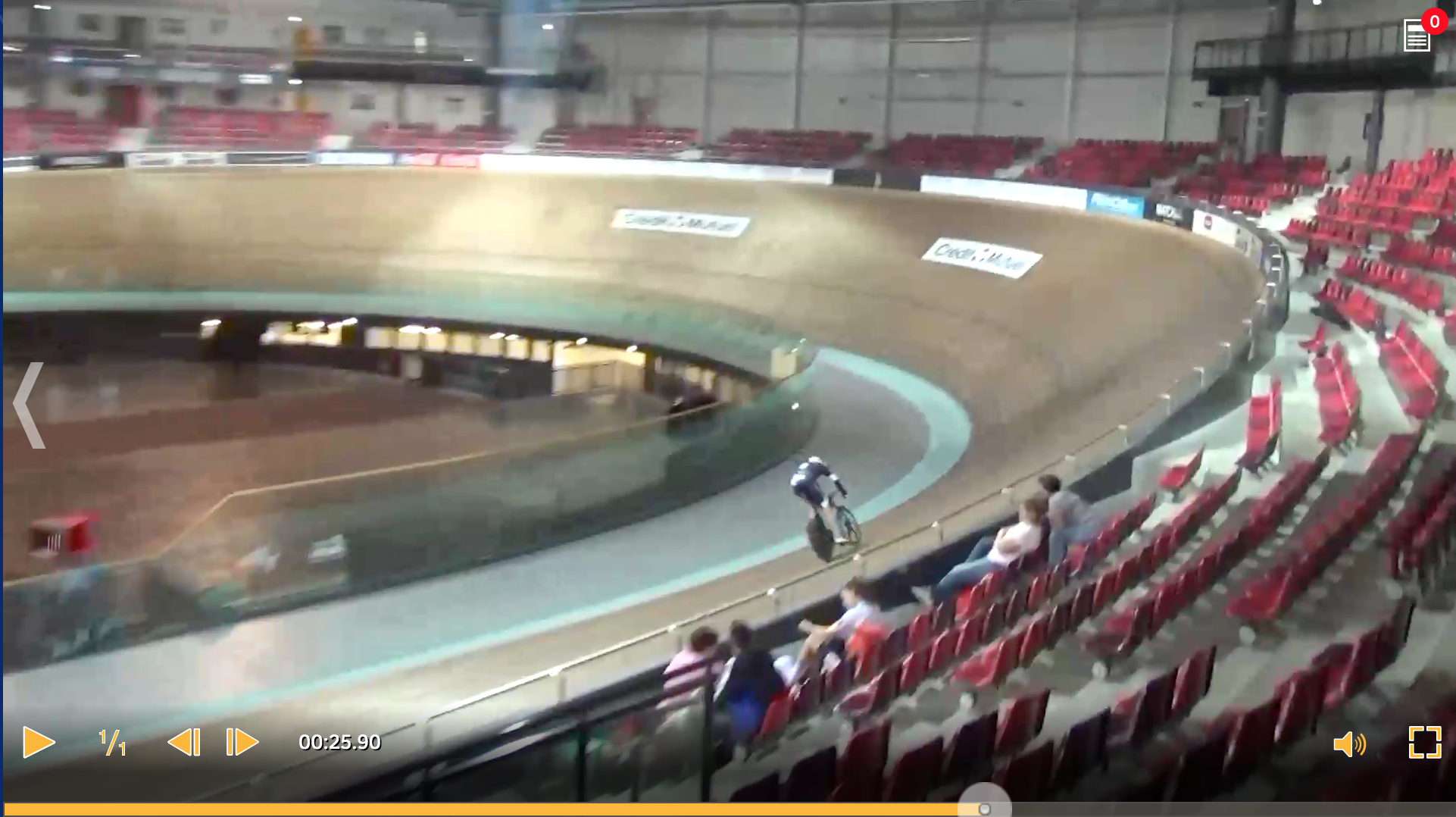}
\put (5,45) {\large \color{white} c)}
\put (-5,-15) {\large e)}
\end{overpic}};
\end{tikzpicture}
\hspace{0.5cm}
\begin{tikzpicture}[scale=0.5]
\node at (0,0) {\begin{overpic}[width=0.33\textwidth,trim={8cm 4cm 8cm 4cm},clip]{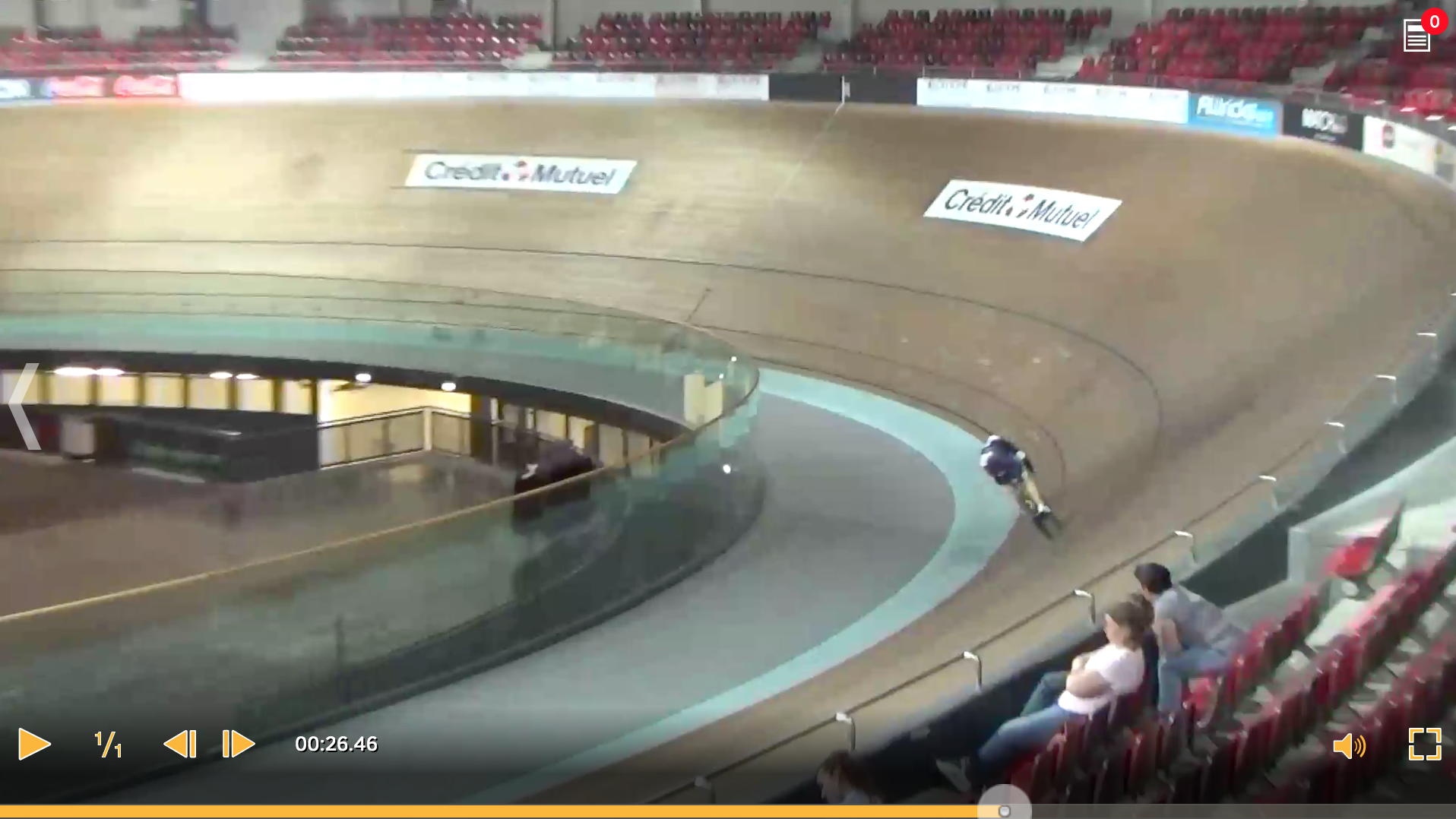}
\put (5,45) {\large \color{white} d)}
\end{overpic}};
\draw[line width=1,->] (1,-1.2)--(1.5,-0.2);
\node at (1.6,-1.5) {\bf c\^{o}te d'azur};
\draw[line width=1,->] (2,1.0)--(1.5,0.5);
\node at (1.6,1.5) {\bf black line};
\end{tikzpicture}\\
\vspace{0.5cm}
\begin{tabular}{|c|c|c|c|c|}
\hline
Pos. & Athlete & Country & Time (s) & $\overline{v}$ (km/hr) \\
\hline
1 & Jason Kenny & GBR & 9.551 & 75.384 \\
2 & Callum Skinner & GBR & 9.703 & 74.203 \\
3 & Matthew Glaetzer & AUS & 9.704 & 74.196 \\
4 & Denis Dmitriev & RUS & 9.774 & 73.664 \\
5 & Gr\'{e}gory Baug\'{e} & FRA & 9.807 & 73.416\\
\hline
\end{tabular}
\caption{(a,b,c,d) Snapshots at different times during a descent trajectory in a velodrome qualification time trial (b,c,d correspond to $0.42,0.84,1.28$ s after the descent). (e) Top five track times taken from the qualification round of the Rio 2016 Olympic games \cite{url1}.\label{snaps}} 
\end{figure}
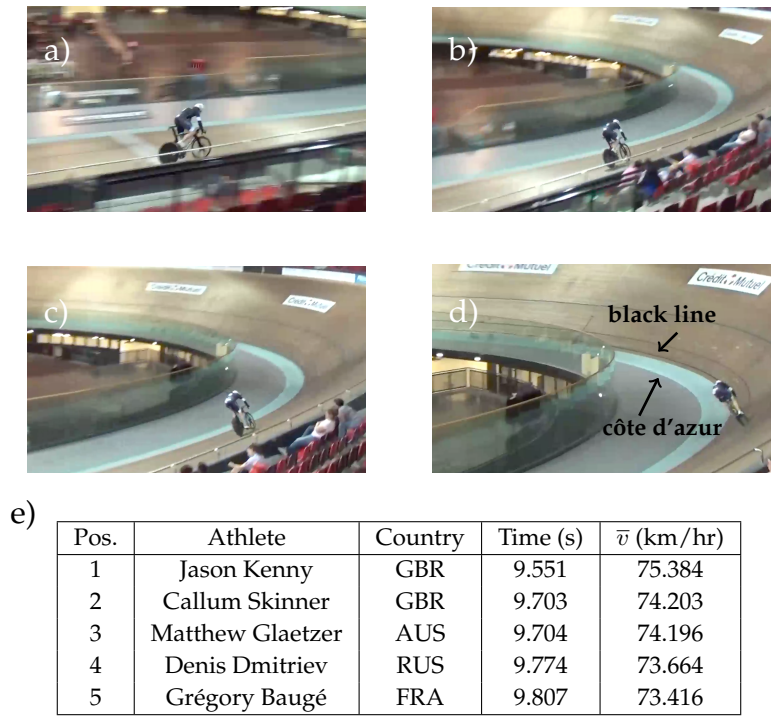

In sports, a natural extension of the Brachistochrone problem is the motion of a cyclist in a velodrome. Track cyclists compete to move around the sloped velodrome surface as fast as possible, much like a higher dimensional Brachistochrone.
Whilst there are many different types of velodrome races, the type of race that lends itself most obviously is the qualification time trial \cite{dorel2005torque}. In this case cyclists complete three and a half laps of the velodrome, where only the time for the final $200\un{m}$ of the last lap is measured. The cyclists build up speed on a single high gear roughly over the first two laps \cite{faria2005science}, staying as high up on the velodrome slope as possible to maintain large potential energy. Then, on the final lap they descend the slope and sprint around the track as fast as they possibly can. 
Similarly to the Brachistochrone, the choice of the descent trajectory is critical. Time is lost if the descent trajectory is too steep or too shallow. The optimal trajectory must balance the exchange between potential and kinetic energy perfectly.

In Fig.  \ref{snaps}(a-d) we show four early time snapshots taken from a velodrome qualification race at the velodrome of Montigny le Bretonneux, France. As the cyclist enters the final lap (at an initial speed of around 58 km/hr), they descend into the sharp corner of the velodrome over a period of around 1 s. 
After the descent, cyclists typically remain at the \textit{black line}, which is at the bottom of the slope, within around $1\un{m}$ of the blue \textit{c\^{o}te d'azur} lane (see Fig.  \ref{snaps}(d)), for the duration of the final lap. Hence, the initial descent is of critical importance because it is the only period of the race where variation is observed between different cyclists. In Fig.  \ref{snaps}(e) we also display the final times of the top five cyclists from the Rio 2016 Olympic games. The difference in time between cyclists is usually of the order of one tenth, and sometimes one hundredth of a second (corresponding to $0.1-1\%$ of the total time). Hence, if the track time can be reduced even slightly by choosing a better descent trajectory, this can have a significant impact on the final ordering of the athletes.

In this study we show how to find the optimum trajectory of a cyclist in a velodrome by modifying the classical formulation of the Brachistochrone problem, treating the cyclist as an active particle on a surface. Since the velodrome track is naturally decomposed into straight and curving sections, we start by studying two canonical cases of motion of a cyclist on a plane and a cone, for which analytical solutions can be found using the Euler-Lagrange equations. Then, using geometrical data taken from the velodrome of Montigny le Bretonneux, we reconstruct the velodrome surface and solve the corresponding optimisation problem using a numerical method that extends from the previous examples. After validating the model by comparison against cyclist velocity and power data, we discuss the various parameters of the problem, as well as the effects of fatigue.


\section{Brachistochrone on a plane.}

Before discussing motion on two-dimensional surfaces, let us first summarise the classic Brachistochrone problem, formulated in the Euler-Lagrange setting \cite{hand1998analytical,butkov1968mathematical}.

Consider a particle of mass $m$ that moves in the vertical plane $(x,z)$ under gravity $g$. This is equivalent to motion on a two-dimensional plane in the case where the plane makes an angle $\alpha=\pi/2$ with the horizontal, where $\alpha$ is illustrated in Fig.  \ref{planecone}(a). We seek to minimise the total time for the particle to move along a trajectory from position $(0,0)$ to $(L,-H)$, which is given by
\beq
T=\int_0^{s_0} \frac{1}{v(s)}\,\mathrm{d}s,\label{time}
\eeq
where $s$ is the arclength of the trajectory, varying from $0$ to $s_0$, and $v=\sqrt{\dot{x}^2+\dot{z}^2}$ is the speed of the particle. 
Neglecting friction, the total energy of the particle is conserved, such that
\beq
\frac{1}{2}mv^2+mgz =0,\label{energy}
\eeq
where we have assumed that the particle is initially at rest. Hence, by using (\ref{energy}) and by rewriting (\ref{time}) in terms of $x$ and $z$, where $\mathrm{d}s=\mathrm{d}x\sqrt{1+(\mathrm{d}z/\mathrm{d}x)^2}$, the total time is
\beq
T=\frac{1}{\sqrt{2g}}\int_0^L \sqrt{\frac{1+(\mathrm{d}z/\mathrm{d}x)^2}{-z}}\,\mathrm{d}x.\label{time2}
\eeq
This quantity (\ref{time2}) can be minimised by solving the Euler-Lagrange equation for the function $z(x)$, which is
\beq
2z\frac{\mathrm{d}^2 z}{\mathrm{d}x^2}+\lb \frac{\mathrm{d}z}{\mathrm{d}x}\rb^2+1=0,\label{bchone}
\eeq
together with the boundary conditions
\begin{align}
z(0)&=0,\\
z(L)&=-H.\label{bc2}
\end{align}
In the case where the final height $-H$ is included as an optimisation variable, the boundary condition (\ref{bc2}) is replaced by
\beq
\frac{\mathrm{d}z}{\mathrm{d}x}(L)=0.
\eeq
It is well-known that the solution is a cycloid which, in the case of the latter boundary condition, is given parametrically by
\beq
x(\lambda)=\frac{L}{\pi}(\lambda-\sin \lambda),\quad z(\lambda)=-\frac{L}{\pi}(1-\cos \lambda),\label{param2}
\eeq
for $\lambda\in[0,\pi]$.

To extend the above formulation to motion on a sloping plane is relatively straightforward. Let us now consider a coordinate system $(x,y,z)$, in which the particle moves on a plane $z=y \tan\alpha$ that makes a constant angle $\alpha$ with the horizontal. By considering the rotated coordinate $Y=y/\cos\alpha$ that lies in the plane, it follows that the total energy is given by
\beq
\frac{1}{2}m\lb\dot{x}^2+\dot{Y}^2\rb+mg Y\sin \alpha =0.
\eeq
Hence, the total time for the descent is
\beq
T=\frac{1}{\sqrt{2g\sin\alpha}}\int_0^L \sqrt{\frac{1+(\mathrm{d}Y/\mathrm{d}x)^2}{-Y}}\,\mathrm{d}x,
\eeq
which is equivalent to the classic Brachistochrone but with a modified gravity $g'=g\sin\alpha$, and a trajectory $Y(x)$ that lies in the plane. Hence, the parametric solution is the same as before (\ref{param2}), except with $z$ replaced by $Y$, and a total descent time multiplied by a factor of $1/\sqrt{\sin\alpha}$.

\section{Brachisto-`cone'}

The next most canonical case of a two-dimensional surface is a cone. This is of particular interest to track cycling because the sharp corner at each end of the velodrome is approximately conical, as we discuss later. By converting the above formulation to cylindrical polar coordinates, it is possible to write down the energy equation for motion on the surface of a cone, which is
\beq
 \frac{1}{2}mv^2+mg(R-R_0)\sin \alpha = 0,
\eeq
where $R=r/\cos\alpha$ is the rotated radial coordinate, and $R_0$ is the initial position of the cyclist. Unlike the planar case, where the initial position is irrelevant, in the conical case $R_0$ is a necessary parameter, and is related to the curvature at the initial position. The resulting time-minimisation problem is written in terms of the integral 
\beq
T=\frac{1}{\sqrt{2g\sin\alpha}}\int_{\theta_0}^{\theta_1} \sqrt{\frac{R^2+(\mathrm{d}R/\mathrm{d}\theta)^2}{R_0-R}}\,\mathrm{d}\theta,
\eeq
where $\theta_1-\theta_0$ is the angle traced out by the trajectory. The resulting Euler-Lagrange equation for the trajectory $R(\theta)$ is
\beq
\frac{\mathrm{d}^2R}{\mathrm{d}\theta^2}+\frac{3R-4 R_0}{2(R_0 -R)R}\lb\frac{\mathrm{d}R}{\mathrm{d}\theta}\rb^2 + \frac{(R -2R_0)R}{2(R_0 -R)}=0,\label{bcone}
\eeq
with boundary conditions
\begin{align}
R(\theta_0)&=R_0,\\
\frac{\mathrm{d}R}{\mathrm{d}\theta}(\theta_1)&=0.\label{rbc2}
\end{align}
The Brachisto-`cone' problem (\ref{bcone}) is different from the Brachistochrone problem (\ref{bchone}) since it also takes into account the effects of rotation, such as the centrifugal force. Such forces play an important role in track cycling, because they allow the cyclist to perform tight corners at high velocity by tilting their bike dramatically in the direction of the bend.

Owing to the more complicated form of (\ref{bcone}) neither an explicit nor a parameterised closed form solution is available (though one constant of integration can be found by considering the Beltrami identity). However, we can solve the boundary value problem (\ref{bcone})-(\ref{rbc2}) numerically. In the subsequent sections, we refer to the solutions in the two cases of the planar Brachistochrone and the Brachisto-`cone' as \textit{analytical}, but indeed only the planar case is in closed form.

\section{Cycling Brachistochrone}

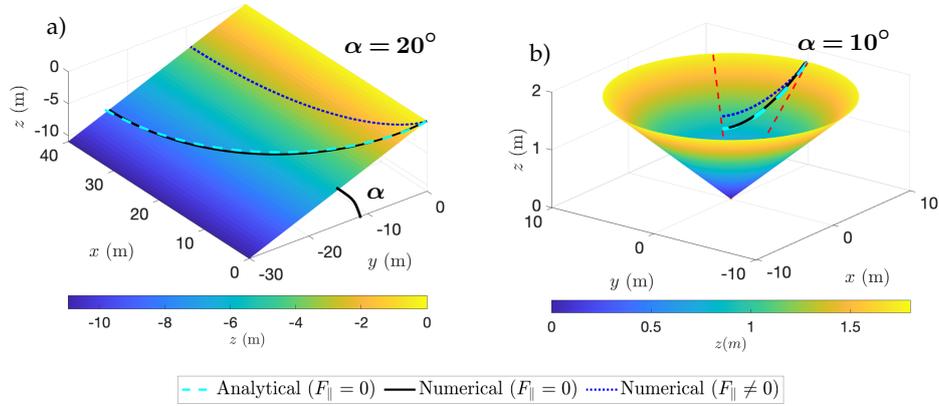
\begin{figure}
\centering
\begin{tikzpicture}[scale=0.4]
\node at (0,0) {\includegraphics[width=0.45\textwidth]{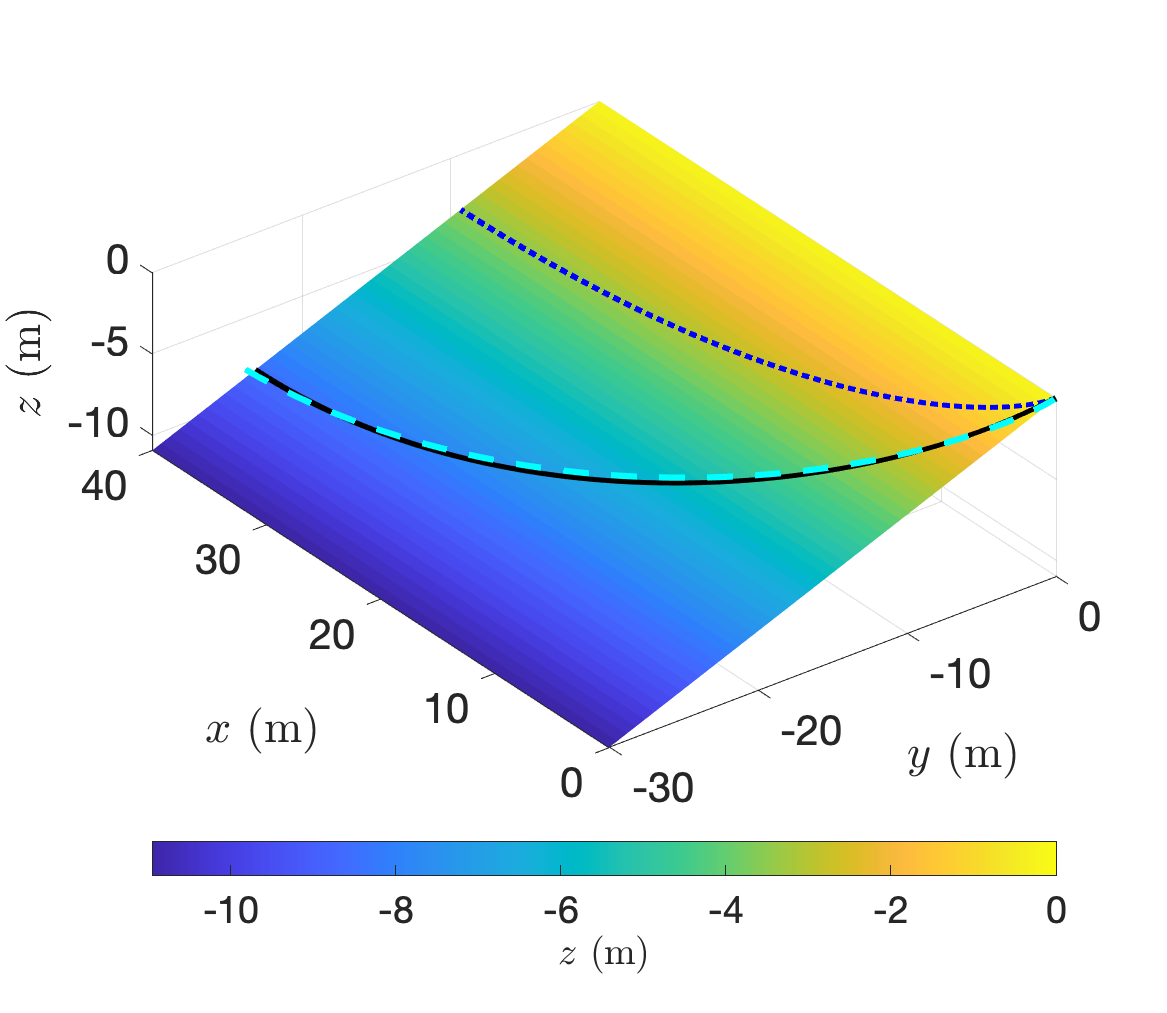}};
\node at (-6,4.5) {a)};
\node at (5,4) {$\boldsymbol{\alpha=20^\circ}$};
\node at (4.5,-1) {$\boldsymbol{\alpha}$};
\draw[line width=1]  (4,-1.8) .. controls (3.8,-1.2) .. (3.2,-0.8);
\end{tikzpicture}
\begin{tikzpicture}[scale=0.4]
\node at (0,0) {\includegraphics[width=0.45\textwidth]{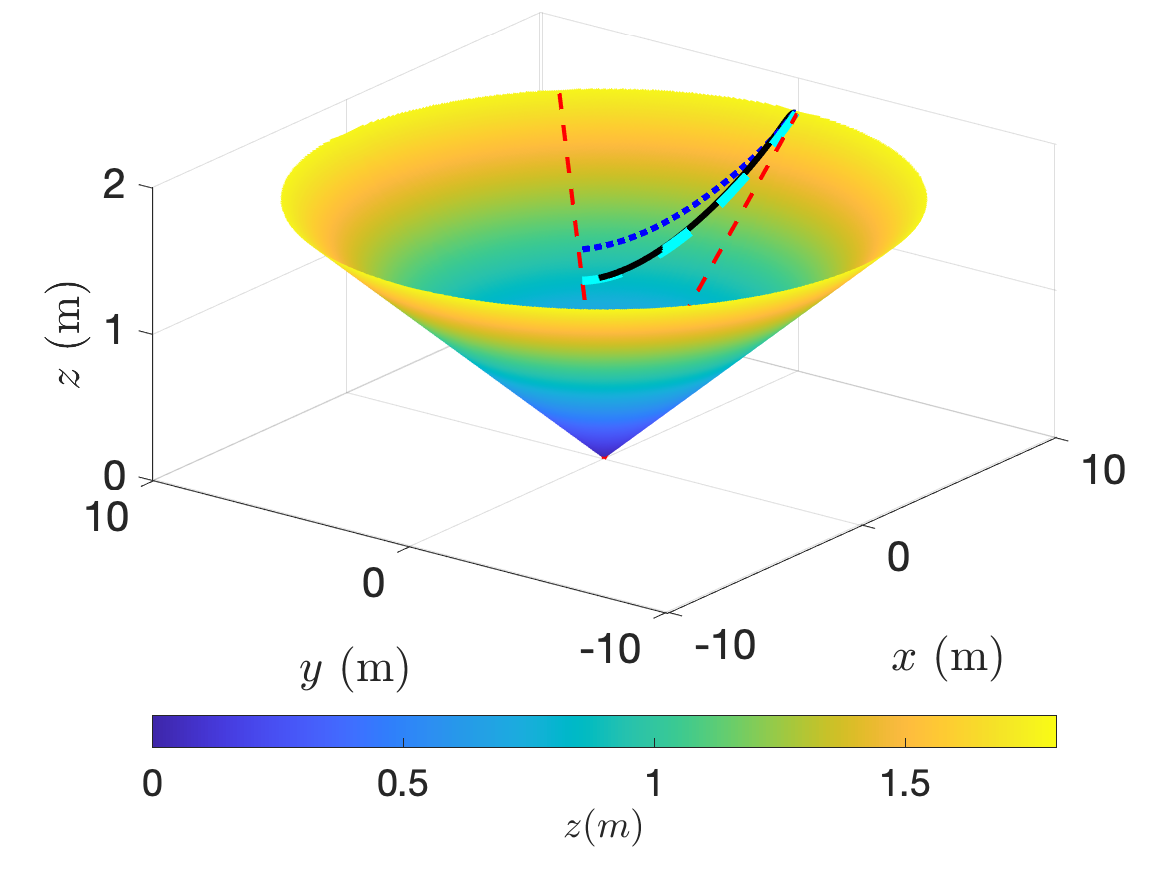}};
\node at (-6,4.5) {b)};
\node at (4,5) {$\boldsymbol{\alpha=10^\circ}$};
\end{tikzpicture}\\
\includegraphics[width=0.6\textwidth]{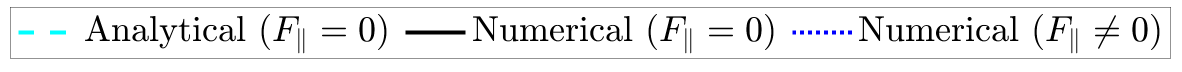}
\caption{Cycling on the planar Brachistochrone (a) and the Brachisto-`cone' (b). In the planar case, we choose a slope angle of $\alpha=20^\circ$ and a total distance of $L=40\un{m}$. In the cone case we take $\alpha=10^\circ$, a starting radius of $R_0\cos\alpha=10\un{m}$ and a total rotation of $\theta_1-\theta_0=\pi/4$. \label{planecone}}
\end{figure}

Let us replace the particle in the above examples by a \textit{point cyclist} (or an active particle). In this case, the cyclist applies a pedal thrust in the direction of motion and experiences aerodynamic drag, such that the total energy is no longer constant.
Thus, the dynamics of the cyclist in the planar case, which we derive in Appendix \ref{appA}, are given by
\begin{align}
 m\ddot{x}&=\frac{\dot{x}F_\parallel}{v}-\frac{ \dot{y} F_\perp}{
 v\cos\alpha},\label{dyn1}\\
 m\ddot{y}&=-mg\sin\alpha\cos\alpha+\frac{\dot{y}F_\parallel}{v}+\frac{  \dot{x}\cos\alpha F_\perp}{
 v}.\label{dyn2}
\end{align}
where $F_{\perp}(t)$ is the leaning force perpendicular to the direction of motion (which does not contribute to the energy), and $F_{\parallel}(t)$ is the force parallel to the direction of motion, which is divided into a pedal thrust and a drag force $F_{\parallel}=F_{p}(t) - F_d(t)$. We model the drag using the parameterisation $F_d=1/2\rho C_d Av^2$, where $C_d$ is the drag coefficient, $A$ is the combined surface area of the cyclist and the bicycle, and $\rho$ is the density of air \cite{mariot1984mechanics,underwood2011aerodynamic,crouch2014flow,crouch2017riding}. 
Note, the only contribution from the normal force is the gravity term in (\ref{dyn2}).
The total energy of the system satisfies
\beq
\frac{\mathrm{d}}{\mathrm{d}t}\lb \frac{1}{2}mv^2+mg Y\sin \alpha \rb=F_{\parallel}v.\label{eloss2}
\eeq
Clearly, if we set $F_{\parallel}=0$ then (\ref{eloss2}) leads to the former Euler-Lagrange formulation. 
If $F_{\parallel}\neq0$ then we cannot solve the problem analytically, but a numerical solution can be found which we discuss shortly.  For the case of the Brachisto-`cone' with forcing, the dynamics are given by (\ref{dynredcone1})-(\ref{dynredcone2}) in Appendix \ref{appA}, and the energy equation is identical to (\ref{eloss2}), except with $Y$ replaced by $R$. Before discussing the solutions to these cases, we first need a model for the pedalling force, and hence the cyclist physiology.

\section{Cyclist physiology}

To model the pedalling force $F_p(t)$, there are certain mechanical and physical considerations that must be taken into account. In particular, since track cyclists must choose a fixed gear ratio for the duration of the race, the pedalling force depends strongly on the instantaneous pedalling rate, and this relationship depends on the physiology of the individual cyclist.

As shown by Dorel et al. \cite{dorel2005torque}, the pedal torque that a cyclist applies in a sprint is a linearly decreasing function of the pedalling frequency. This linear relationship is characterised by two coefficients $T_{max}$ and $\omega_{max}$, which correspond to the maximum possible torque (occurring at zero pedalling frequency) and the maximum possible pedalling frequency (occurring at zero torque). Each cyclist has a sprint performance characterised by these two parameters, and these are easily measured with a pedalling experiment.

The pedalling torque and frequency are related to the pedalling force $F_p$ and speed $v$ via the development $D$, which is the distance travelled by one single rotation of the pedals (analogous to the gear). Hence, the pedalling force is given by the linear relationship
\beq
F_{p}=\frac{2\pi T_{max}}{D}\lb 1 - \frac{2\pi v}{D\omega_{max}} \rb.\label{hill}
\eeq
This relationship is similar to the force-velocity equation related to muscle physiology, sometimes called the Hill equation \cite{hill1938heat}. Note that (\ref{hill}) is only valid for pure anaerobic respiration, and does not include the effects of fatigue. We will discuss the effects of fatigue later in Section \ref{fatsec}.

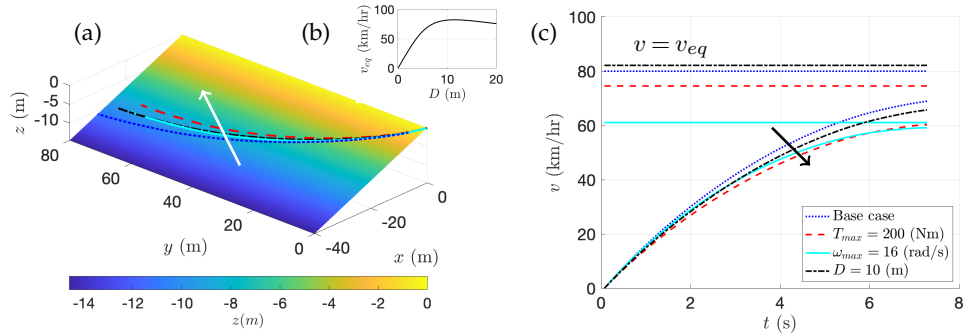
\begin{figure}
\centering
\begin{tikzpicture}[scale=0.5]
\node at (0,0) {\includegraphics[width=0.45\textwidth]{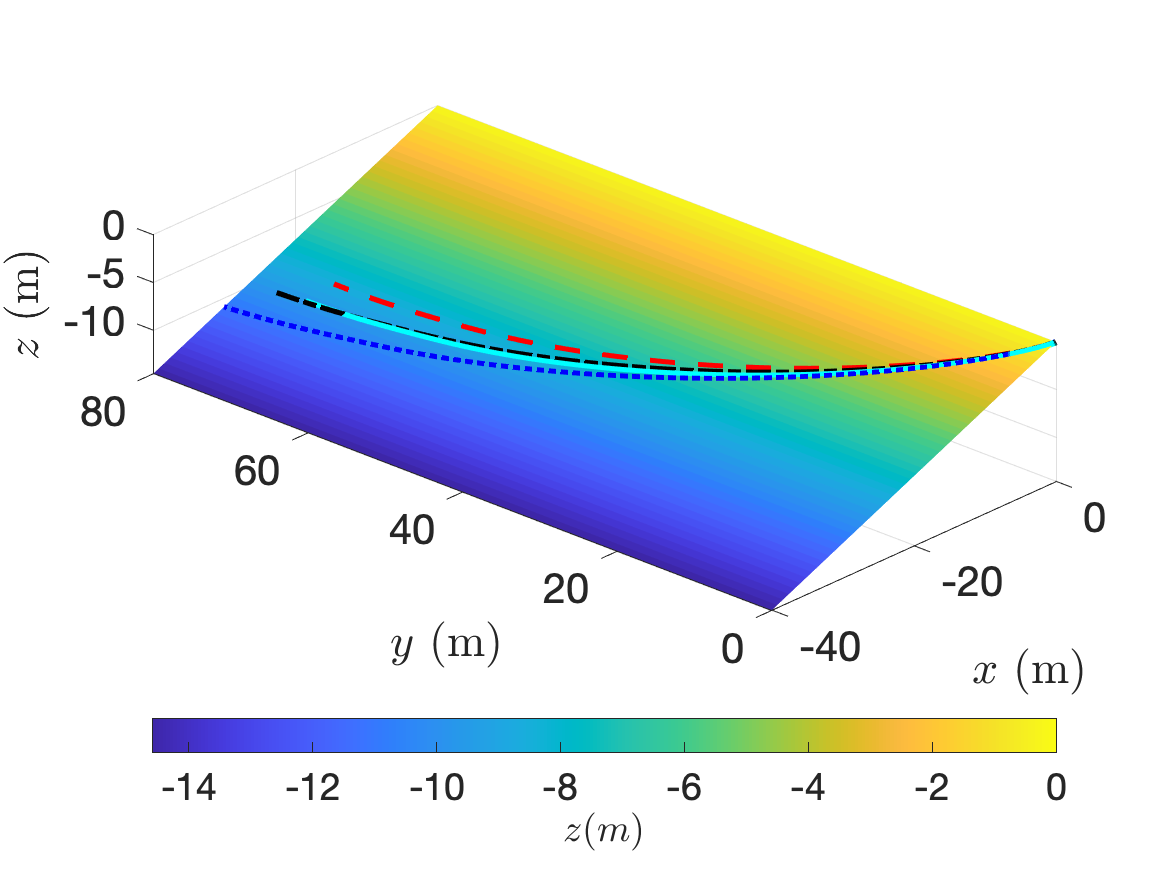}};
\draw[line width=1,white,->] (0,0) -- (-1,2);
\node at (5.1,3) {\includegraphics[width=0.15\textwidth]{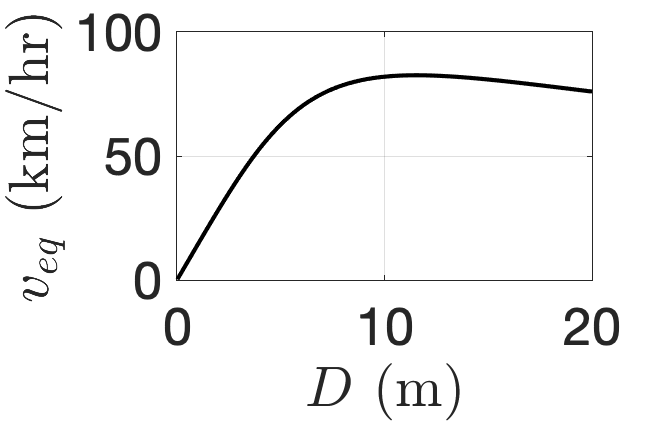}};
\node at (14,0) {\begin{overpic}[width=0.45\textwidth]{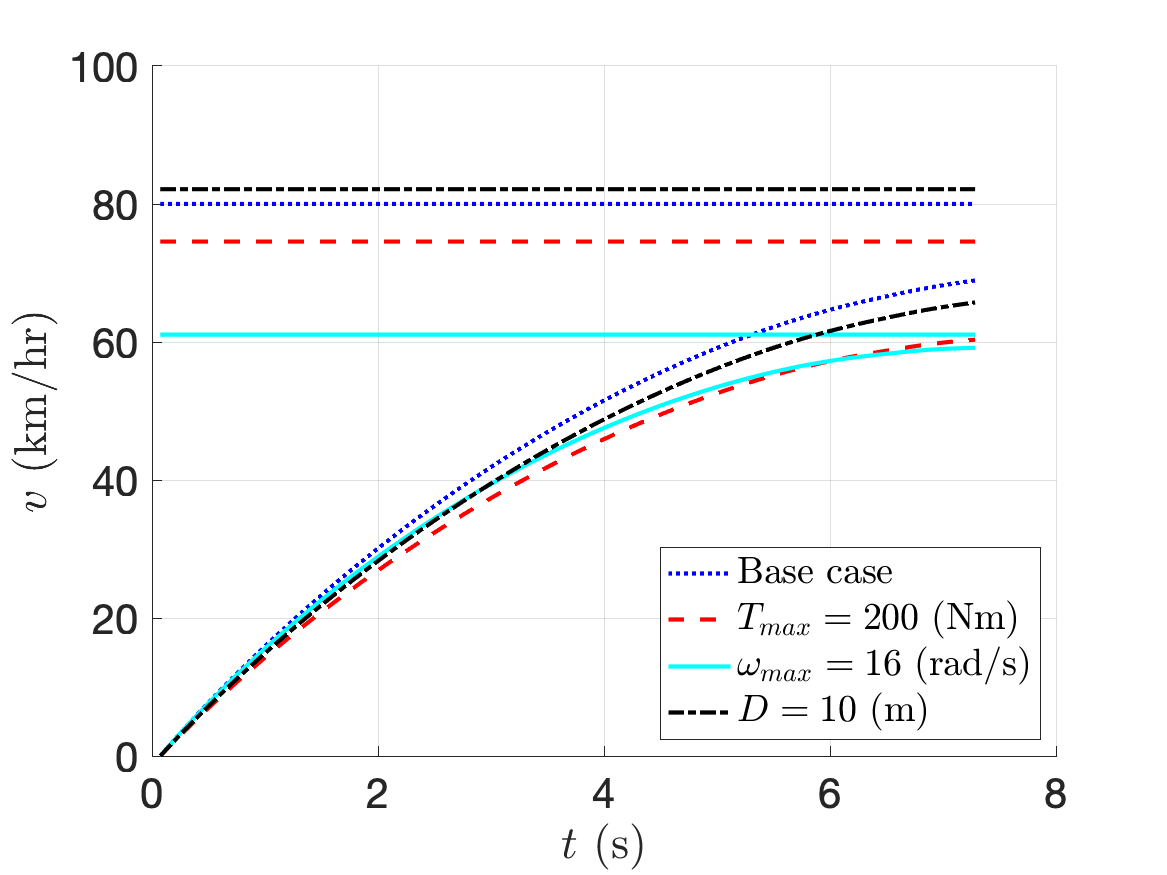}
\put (20,62) {\color{black} $v=v_{eq}$}
\end{overpic}};
\node at (-4,3.5) {(a)};
\node at (2,3.5) {(b)};
\node at (8,3.5) {(c)};
\draw[line width=1,->] (14,1) -- (15,0);
\end{tikzpicture}

\caption{Study of the cycling physiology parameters. In the base case, we have $T_{max}=260\un{Nm}$, $\omega_{max}=25\un{rad/s}$, and $D=8.5\un{m}$. In each of the other cases, we vary just one of these parameters and plot the trajectory (a) and the velocity (c). The equilibrium velocity \eqref{veq} for each case is indicated in (c), and the dependence on $D$ is shown in the insert (b).
\label{cyclist_params}}
\end{figure}

In Fig.  \ref{planecone} we plot solutions to both the planar Brachistochrone and the Brachisto-`cone' problem. Light blue dashed curves correspond to the analytical solution in the case of zero pedalling and drag force $F_{\parallel}=0$. Black curves correspond to the numerical solution to the equivalent optimal control problem, which is achieved by formulating an interior point constrained optimisation \cite{nocedal2006numerical,wachter2006implementation} using the dynamics (\ref{dyn1})-(\ref{dyn2}) to govern the variables $x(t),y(t)$, and using the forcing $F_{\perp}(t)$ as a control function (see Appendix \ref{appB}). The analytical solution is useful for validating the numerical approach, giving us confidence when applying it to the case of non-zero pedalling and drag force, for which an analytical solution is not available. Such solutions (for $F_{\parallel}\neq0$) are shown on the same plot with dotted blue curves.

To calculate these trajectories we choose values for the model parameters that correspond to realistic cycling scenarios. We take the combined mass of the cyclist and the bike as $m=86\un{kg}$, and the physiological characteristics $\omega_{max}=25\un{rad/s}$, $T_{max}=260\un{Nm}$, correspond to data taken from an elite athlete \cite{dorel2005torque}. The product of the drag coefficient and the surface area is $C_dA=0.22\un{m^2}$, which is equivalent to a streamlined cycling position and modern equipment \cite{martin1998validation,lukes2005understanding,underwood2011aerodynamic,crouch2014flow,chowdhury2014experimental,crouch2017riding}. Finally, we choose a development of $D=8.5\un{m}$, which is typical for velodrome sprint qualification trials.

In Fig. \ref{cyclist_params}, we illustrate how each of these parameters affects cyclist performance. We display the trajectory and velocity profile for the planar cycling Brachistochrone (from Fig. \ref{planecone}(a)) which we use as a base case (though extended to an $80\un{m}$ track), as well as three other curves for which we perturb each of the three physiological parameters $T_{max}$, $\omega_{max}$ and $D$. In each case, the perpendicular force profile $F_\perp(t)$ and total time are fixed at their base case values. 
The perturbed values of each of these parameters ($T_{max}=200\un{Nm}$, $\omega_{max}=16\un{rad/s}$, $D=10\un{m}$) result in a reduced velocity and a shallower descent trajectory. In the case of $T_{max}$ and $\omega_{max}$ this can be explained by a reduction in the forward thrust and maximum velocity, respectively. The effect of increasing the development $D$ is less obvious, since this simultaneously increases the maximum velocity whilst decreasing the maximum thrust. However, this becomes more clear by considering the equilibrium velocity
\beq
v_{eq}=\frac{ -4 \pi^2 T_{max} + 
    2\sqrt{\pi T_{max}(
    4 \pi^3 T_{max} + \rho C_d A D^3  \omega_{max}^2)}}{\rho C_d A D^2  \omega_{max}},\label{veq}
\eeq
which is equivalent to a balance between the pedalling force and the drag force. 
As indicated by Fig. \ref{cyclist_params}(b), $v_{eq}$ is non-monotonic function of $D$ for typical parameter values, and has a maximum at $D\approx11.5\un{m}$. Therefore, whilst increasing $D$ from $8.5\un{m}$ to $10\un{m}$ raises the equilibrium velocity, the pedal force is simultaneously reduced, resulting in a longer time to achieve equilibrium, and hence a slower, shallower trajectory.

\section{From Brachistochrone to velodrome}
\label{fatsec}
\subsection{Validation}

\begin{figure}
\centering
\begin{tikzpicture}[scale=0.5]
\node at (0,0) {\includegraphics[width=0.5\textwidth]{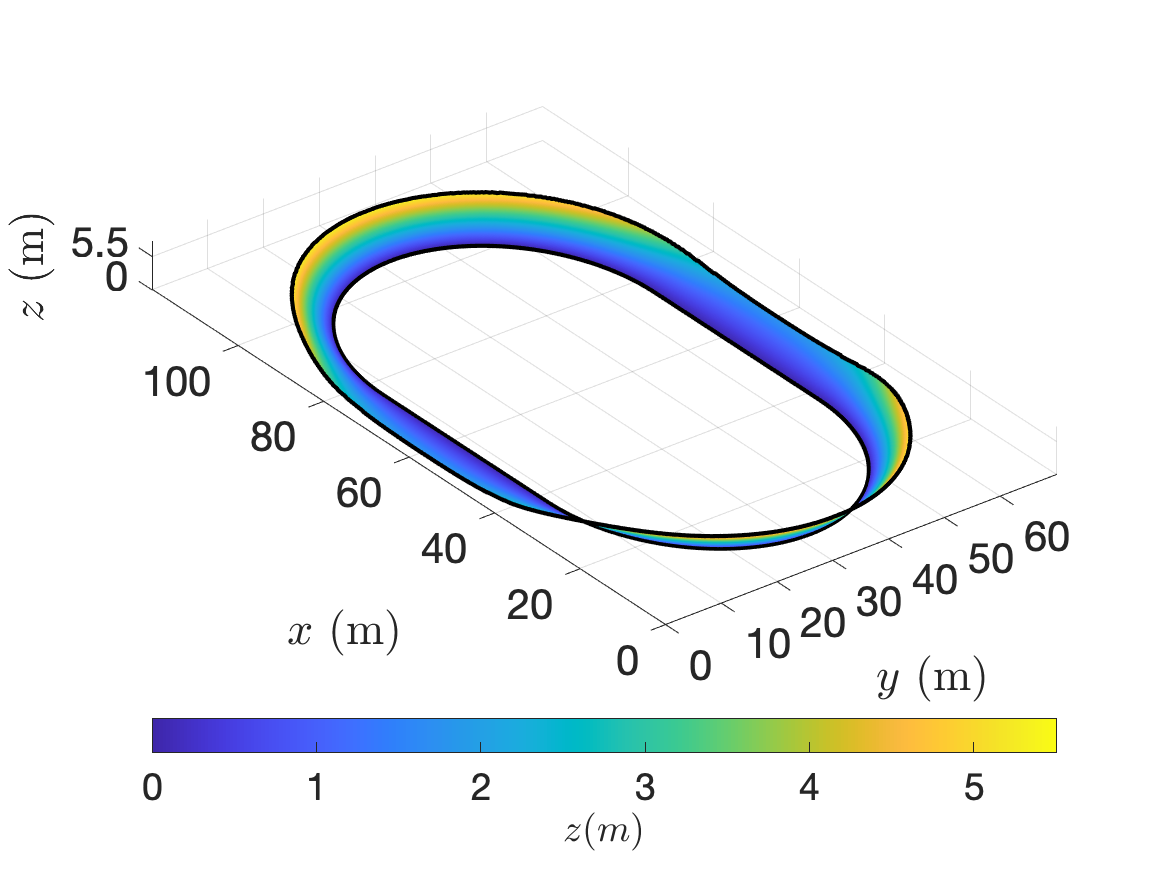}};
\draw[line width=1,dotted] (-2.5,0.6) -- (0.7,1.8);
\draw[line width=1,dotted] (-0.6,-0.7) -- (2.6,0.5);
\draw[line width=0.5,<->] (2.3,0.5) -- (0.6,1.6);
\node at (1.1,0.8) {$L$};
\node at (-1.8,1.4) {$a$};
\node at (-0.7,1.9) {$b$};
\node at (2.2,2.5) {$s=0$};
\draw[line width=0.5,<->,] (-1.2,1.3) -- (-2.1,1.9);
\draw[line width=0.5,<->,] (-1.1,1.4) -- (0.2,1.9);
\draw[line width=0.5,->]  (1.3,2.5) .. controls (0.5,3) .. (0,3);
\draw[line width=1.5,white, dotted] (1.6,1.9) -- (1.0,1.5);
\node at (-5,3) {a)};
\end{tikzpicture}
\begin{tikzpicture}[scale=0.4]
\draw [line width=1, black] (3,-1)--(0,0);
\draw [line width=1, black] (3,-1)--(0,4);
\draw [line width=1, black] (0,0)--(0,4);
\draw [line width=1, black,dotted] (13,0)--(10,1);
\draw [line width=1, black,dotted,->] (13,0)--(11.5,0.5);
\draw [line width=1, black,dotted,->] (13,0)--(11.5,0);
\draw [line width=1, black] (13,0)--(10,5);
\draw [line width=1, black,dotted] (10,1)--(10,5);
\draw [line width=1, black]  (0,4) parabola bend (5,4.5) (10,5);
\draw [line width=1, black]  (3,-1) parabola bend (7,0) (13,0);
\draw [line width=1, black,dotted]  (0,0) parabola bend (4,1) (10,1);
\node [draw, line width=1, circle]at (5.5,2){} ;
\node [draw, line width=1, circle]at (6.3,2.2){} ;
\draw[line width=1, black] (5.5,2) -- (5.6,2.4) -- (6.1,2.6) -- (6.3,2.2) -- (5.9,2.1) ;
\draw[line width=1, black,->] (5.5,2)--(4.4,2) ;
\draw [line width=1, black] (5.5,2.6)--(5.9,2.1) -- (6.2,2.8);
\draw [line width=1, black] (5.4,2.6)--(5.7,2.6) ;
\draw [line width=1, black] (6.05,2.8)--(6.25,2.8) ;
\node at (4,2.7) {$\boldsymbol{F}(t)$};
\node at (7.2,3) {$\boldsymbol{x}(t)$};
\node at (5,5) {$z=n\tan\alpha(s)$};
\node at (12,-0.5) {$s$};
\node at (12,1) {$n$};
\node at (12.5,3) {\rotatebox{300}{$W=7.9$m}};
\draw [line width=1,<->] (13.5,0)--(10.5,5);
\draw [line width=1,->] (0,0)--(0,2);
\node at (-0.5,2) {$z$};
\node at (-0.5,5) {c)};
\draw [line width=1, black]  (2.4,0) .. controls (2.1,-0.3) .. (2.2,-0.7);
\node at (1.4,0.1) {$\alpha(s)$};
\node at (5,9) {\includegraphics[width=0.4\textwidth]{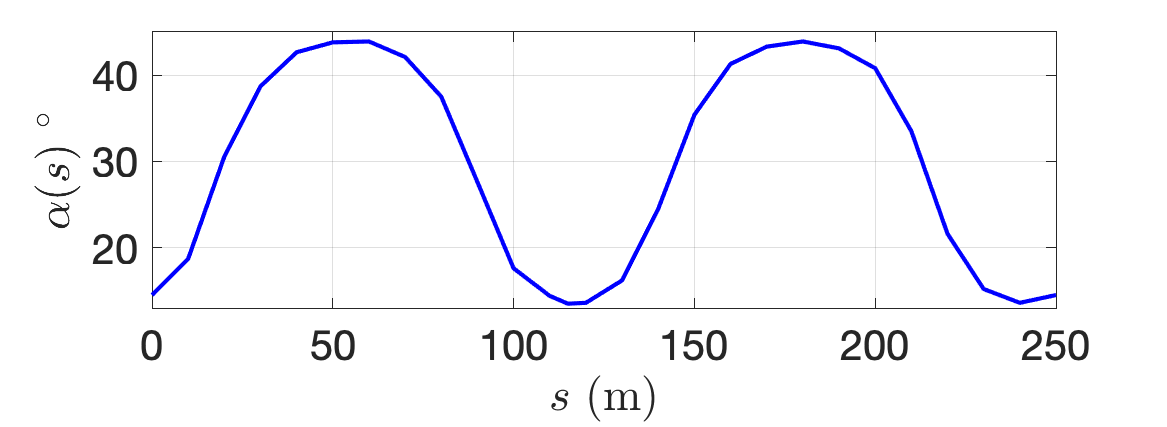}};
\node at (-1.5,11) {b)};
\end{tikzpicture}\\
\caption{(a) Three-dimensional reconstruction of the velodrome of Montigny le Bretonneux. (b)  The velodrome slope angle $\alpha$ measured experimentally as a function of distance around the track $s$. (c) Schematic diagram of the velodrome surface $z=n \tan \alpha(s)$, illustrating the tangent and normal coordinates $(s,n)$, and the cyclist position and force $\boldsymbol{x}(t)$, $\boldsymbol{F}(t)$.   \label{schem}}
\end{figure}
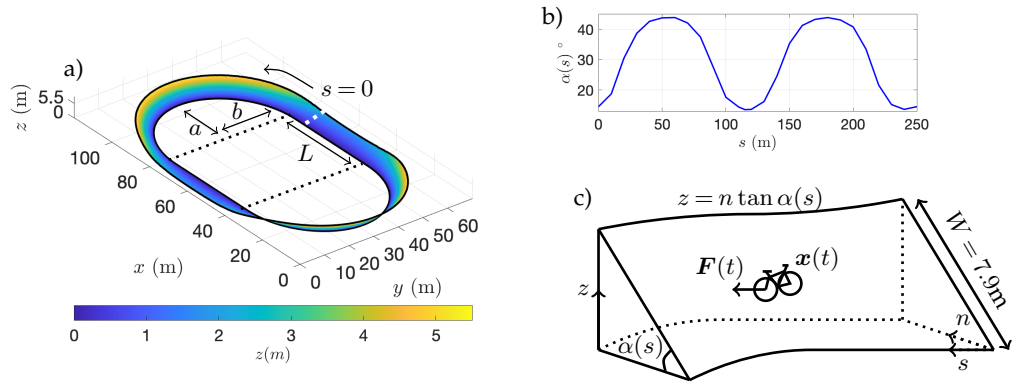

The next step is to apply the above method to find the optimum trajectory on a real velodrome track and validate the model by comparison with cyclist velocity and power data. As a case study, we choose the velodrome of Montigny le Bretonneux in France. For this velodrome, the inside lane is composed of two straight lines of length $L=38\un{m}$, connected by two half-ellipses of semi-major and semi-minor axes $a=29.8\un{m}$ and $b=24.2\un{m}$ (see Fig.  \ref{schem}(a)). Since the velodrome slope varies as one moves around the track, it is convenient to make use of the tangent and normal coordinates measured on the inside lane $(s,n)$ (see the schematic diagram in Fig.  \ref{schem}(c)). In terms of these coordinates, the velodrome surface is written simply as $z=n\tan\alpha(s)$, where the width of the surface is a constant $W=7.9\un{m}$. Therefore, unlike the previous examples, motion on the velodrome is bounded, such that $0\leq n\leq W\cos\alpha$. In Fig.  \ref{schem}(b) we plot the angle $\alpha(s)$, which was measured at the race track using an angle-metre. The slope angle is approximately sinusoidal, varying between $14^\circ$ in the middle of the straight sections to $45^\circ$ in the middle of the curved sections. The reconstructed velodrome surface is shown in (a).


Based on the previous canonical examples, we formulate and solve a numerical optimisation problem for the trajectory, where the dynamics in the straight regions are similar to (\ref{dyn1})-(\ref{dyn2}), and for the elliptical sections of the velodrome we consider a small perturbation from the cone example, where the ellipticity $\epsilon=a/b-1=0.23$ is treated as a small parameter in an an asymptotic expansion (see Appendix \ref{appA}).

In Fig.  \ref{traj} we display the optimal trajectory, as well as plots of velocity $v$ and power $P=vF_p$ as functions of time \cite{martin2007understanding}. For this example, we use the same parameter values as before, except with initial position $s(0)=0,$ $n(0)=W\cos\alpha(0)$, and an initial velocity of $58\un{km/hr}$ purely in the $x$-direction, which corresponds to the typical speed before descent for Olympic athletes (further discussion on optimising the initial conditions of the problem is given in the next section). The optimal trajectory is very similar to the strategy employed by athletes, descending before the first corner and then hugging the inside lane thereafter. The descent follows closely to the position of the cyclist in the snapshots in Fig.  \ref{snaps}(a-d), which we indicate approximately with white stars. The velocity and power in Fig.  \ref{traj}(b,c) are compared to data taken from an elite cyclist in the velodrome of Montigny le Bretonneux over three different race attempts. The close agreement between our model and the data indicates that the cyclist's trajectory is already nearly optimal, and illustrates the validity of our model.
Furthermore, the time to complete the final $200\un{m}$ of the lap is around $9.7\un{s}$, which is close to high-ranking Olympic performance, as seen in the table in Fig.  \ref{snaps}(e).



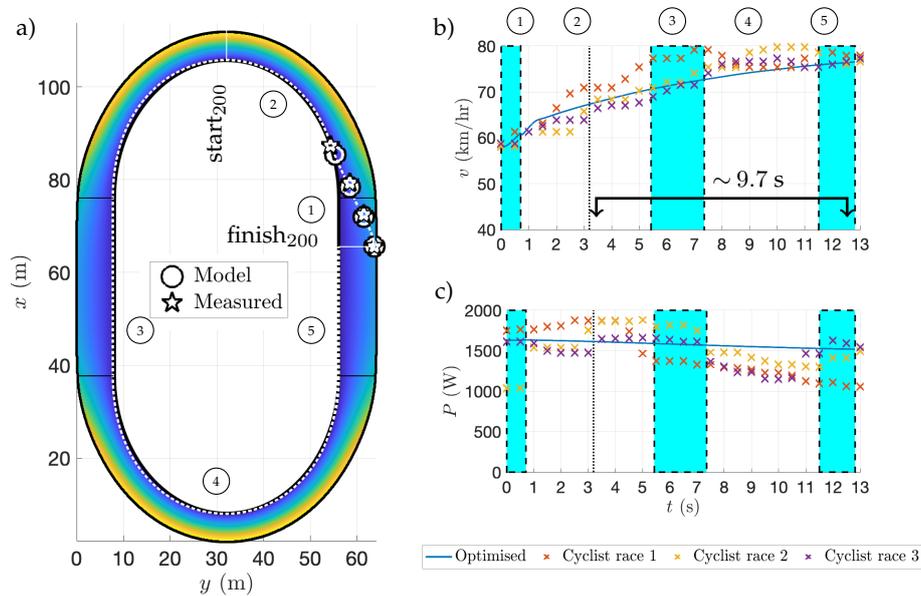
\begin{figure}
\centering
\begin{tikzpicture}[scale=0.5]
\node at (0,1) {\includegraphics[width=0.4\textwidth]{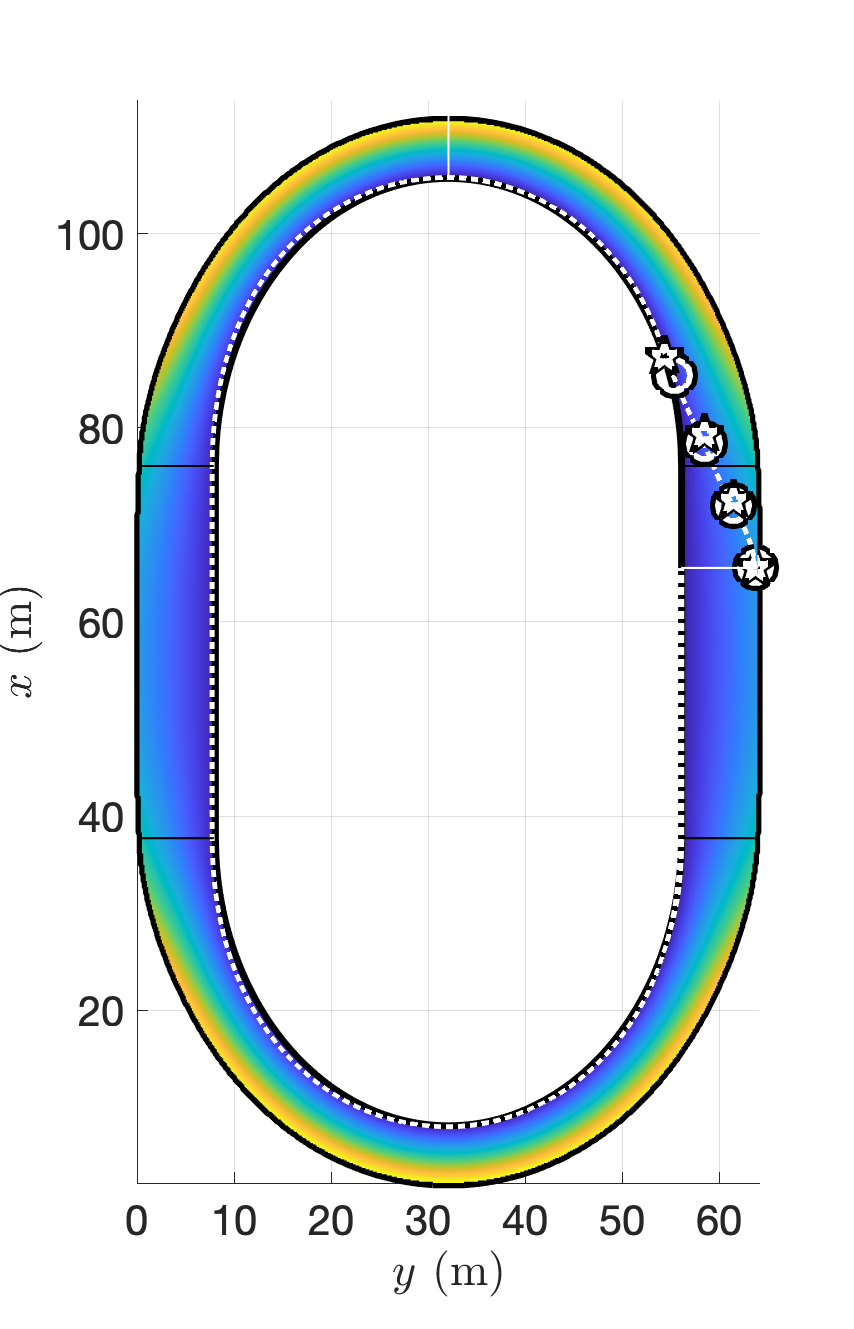}};
\node at (2.5,3.2) {\tiny $\circled{1}$};
\node at (1.5,6) {\tiny $\circled{2}$};
\node at (0,-4) {\tiny $\circled{4}$};
\node at (2.5,0) {\tiny $\circled{5}$};
\node at (1.5,2.5) {finish$_{200}$};
\node at (0,1) {\includegraphics[width=0.15\textwidth]{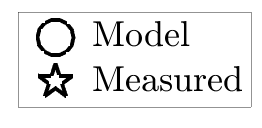}};
\node at (-2,0) {\tiny $\circled{3}$};
\node at (0,5.5) {\rotatebox{90}{start$_{200}$}};
\node at (12,5) {\includegraphics[width=0.45\textwidth]{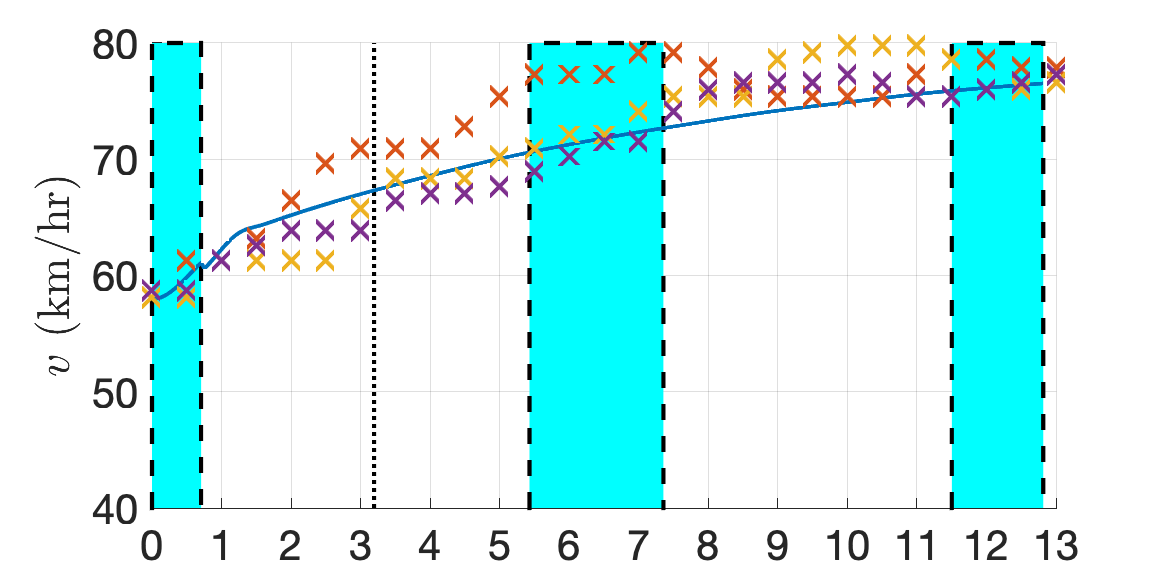}};
\node at (12,-2) {\includegraphics[width=0.45\textwidth]{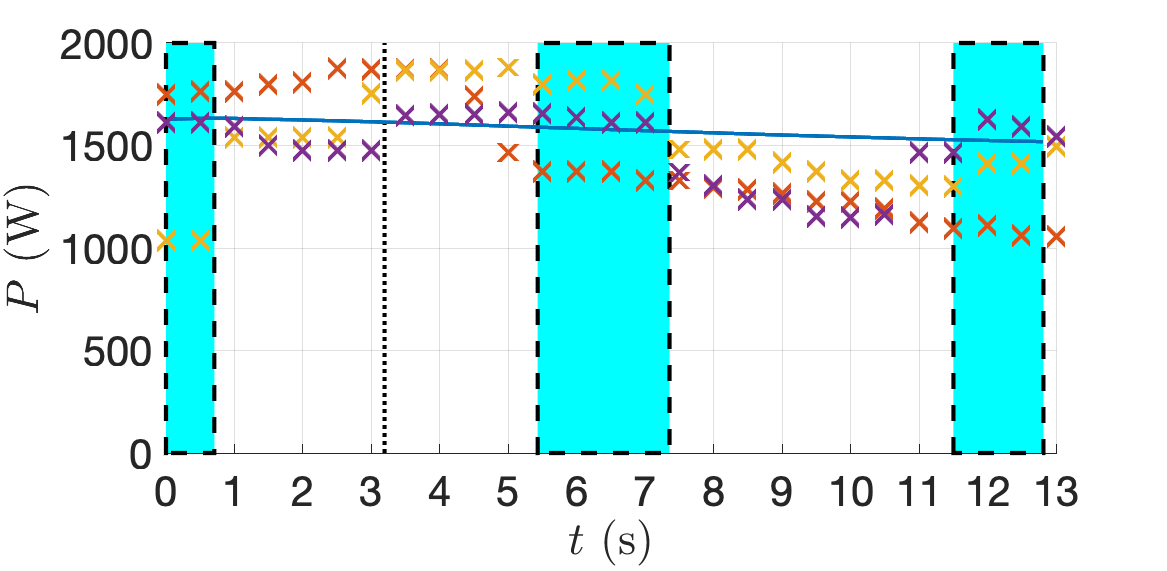}};
\draw[line width=1,<->] (10,3) -- (10,3.5) -- (16.6,3.5) --(16.6,3);
\node at (14,4) {$\sim 9.7$ s};
\node at (8,8.2) {\tiny $\circled{1}$};
\node at (9.5,8.2) {\tiny $\circled{2}$};
\node at (12,8.2) {\tiny $\circled{3}$};
\node at (14,8.2) {\tiny $\circled{4}$};
\node at (16,8.2) {\tiny $\circled{5}$};
\node at (6,8) {b)};
\node at (-5,8) {a)};
\node at (6,1) {c)};
\node at (12,-6) {\includegraphics[width=0.5\textwidth]{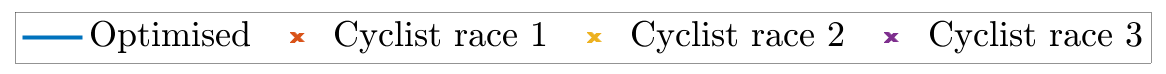}};
\end{tikzpicture}
\caption{(a) Brachistochrone on a velodrome, viewed from above, illustrating the positions that correspond to the snapshots in Fig.  \ref{snaps}(a-d). We also indicate the different curved and straight regions of the velodrome, as well as the start and finish lines for the final $200\un{m}$ of the track. (b,c) Corresponding speed $v$ and power $P$ profiles, compared with measured data from an elite athlete for three different race attempts. Straight regions of the track are illustrated with shading.  \label{traj}}
\end{figure}

\subsection{Discussion of the parameters and the effects of fatigue}



Although we have so far only discussed optimising the cyclist trajectory, there is also the question of the optimal choice of the initial position and velocity of the cyclist. In this section, we discuss these initial conditions, including how fatigue plays a role in their optimal values.

Written in terms of the tangent and normal coordinates, there are four initial conditions $s(0)$, $n(0)$, $\dot{s}(0)$ and $\dot{n}(0)$. 
We find that the optimum value of $n(0)$ is always given by $n(0)=W\cos \alpha(s(0))$, which is in accordance with our observation that track cyclists always begin their descent from the top of the velodrome ramp. This choice of $n(0)$ can be explained by the desire to maximise potential energy before descent. Note, that due to the proximity of the velodrome boundary, this also constrains the initial descent angle of the cyclist, such that $\dot{n}(0)=0$. Hence, the initial speed is simply $v(0)=\dot{s}(0)$.

To determine the remaining optimum parameter values, $s(0)$ and $\dot{s}(0)$, one must consider the effects of fatigue. As discussed by \cite{dorel2005torque}, the relationship (\ref{hill}) is only valid whilst the cyclist sprints using pure anaerobic respiration and, in reality, this can only last for around $\sim5\un{s}$. Afterwards, the effects of fatigue take action, and eventually respiration becomes aerobic. After this point, the power exerted by the cyclist deviates towards a near-constant plateau of around $P_0/m=6.35\un{W/kg}$ \cite{sanders2019anaerobic}. The initial velocity is explained by balancing the pedalling force that corresponds to constant aerobic power ($P=P_0$) and the drag force, such that $P_0/v\sim 1/2\rho C_d A v^2$, which for a cyclist of mass $86\un{kg}$ gives a solution $v=58\un{km/hr}$. If the cyclist were to choose a larger initial velocity than this, they would risk eating up some of their energy budget during the final lap. Hence, they choose to start the descent at the maximum possible velocity without triggering the effects of fatigue prematurely. 

The final parameter of interest $s(0)$ is one of the main differences between cyclist strategies. Cyclists typically start the descent between $s=3\un{m}$ and $s=-11\un{m}$ (where $s$ is measured with respect to the finish line). The descent position $s(0)$ is chosen as a tradeoff between speed and fatigue. If the descent starts too late, then it is not possible to build up enough speed for the final lap. On the other hand, if the descent starts too early then, although the final lap will commence at high velocity, the cyclist will quickly become tired. 

According to \cite{sanders2019anaerobic}, the pedalling power decays exponentially due to fatigue, where the formula
\beq
P=P_0+(P_\mathrm{max}-P_0)e^{-t/\tau},\label{sandersp}
\eeq
gives a good fit with cyclist data using parameter values $P_\mathrm{max}/m=17.4\un{W/kg}$ and $\tau=38\un{s}$. 
The choice of starting position should be such that the cyclist crosses the finish line before fatigue causes any deceleration. Hence, we expect the peak velocity $v_\mathrm{peak}$ to be precisely at the finish line. Therefore, at this moment, we expect a perfect equilibrium between the pedalling power \eqref{sandersp} and the drag power. From this balance, we derive the sprint timescale
\beq
t_\mathrm{sprint}=\tau\log\lb\frac{P_\mathrm{max}-P_0}{1/2\rho C_dAv_\mathrm{peak}^3-P_0}\rb.\label{sprinttime}
\eeq
During the London 2012 Olympics qualification trials, starting positions varied between $s(0)=3\un{m}$ and $s(0)=-11\un{m}$, with corresponding total sprint times (including the descent) of $t_\mathrm{sprint}=13-15\un{s}$. According to \eqref{sprinttime}, these sprint times correspond to peak velocity values of around $v_\mathrm{peak}=74-75\un{km/hr}$, which are close to observations (e.g. $v_\mathrm{peak}=77\un{km/hr}$ in Figure \ref{traj}). Hence, this simple scaling argument provides a good explanation for the choice of starting position. 
To illustrate how the descent position $s(0)$ affects the shape of the descent trajectory, we display different calculated optimum trajectories for various values of $s(0)$ in Appendix \ref{startps}.

\section{Concluding remarks}

Whilst the classic Brachistochrone problem is limited to one dimension, we have shown how a similar formulation can extend the problem to motion on a plane or a cone, for which the Euler-Lagrange equations yield analytical solutions. By extending the classic Brachistochrone using these two canonical cases, as well as including the effects of cyclist physiology, we have shown how to optimise the trajectory of a cyclist in the real example of the velodrome in Montigny le Bretonneux, finding very close agreement with measured athlete data. In addition to the optimum trajectory, we have also discussed optimising the initial conditions before descent, where the effects of fatigue play a role.


This work not only paves the way for future studies of cycling, but also has implications for research across the field of physics on the topic of active particles on surfaces.
For future work, the dynamics of other cyclists in the race could be included \cite{defraeye2014cyclist}, where the effects of slipstreaming and racing psychology must be taken into account \cite{kyle2004aerodynamics}. 

\ack{We thank Tullio Traverso and Benjamin Lallemand for useful discussions. We gratefully acknowledge the support from Ecole Polytechnique for the research program Sciences 2024.}

\appendix

\section{Derivation of the governing equations\label{appA}}

\subsection{Cartesian coordinates}
Consider the motion of a particle on a two-dimensional surface $z=f(x,y)$ under gravity without any other external forcing. The Lagrangian of the system is given by the kinetic energy minus the potential energy
\beq
\mathcal{L}=\frac{1}{2}m\lb \dot{x}^2+\dot{y}^2+\dot{z}^2 \rb - m g z.
\eeq 
Since the motion is confined to the surface, we can eliminate $z$ to give
\beq
\mathcal{L}=\frac{1}{2}m\lb \dot{x}^2\lb 1+\frac{\partial f}{\partial x}^2\rb +\dot{y}^2\lb 1+\frac{\partial f}{\partial y}^2\rb+2\frac{\partial f}{\partial x}\frac{\partial f}{\partial y}\dot{x}\dot{y} \rb - m g f.
\eeq
The Euler-Lagrange equations for this system can be rearranged to give two second order differential equations for $x(t)$ and $y(t)$, which are
\begin{align}
m\ddot{x}&=-\frac{\frac{\partial f}{\partial x} mg + m\frac{\partial f}{\partial x}\lb \frac{\partial^2 f}{\partial x^2} \dot{x}^2 + 2  \frac{\partial^2 f}{\partial x \partial y}  \dot{x}  \dot{y} +   \frac{\partial^2 f}{\partial y^2} \dot{y}^2\rb}{
 1 + \frac{\partial f}{\partial x}^2 + \frac{\partial f}{\partial y}^2},\label{dyn1ap}\\
 m\ddot{y}&=-\frac{\frac{\partial f}{\partial y} mg + m\frac{\partial f}{\partial y}\lb \frac{\partial^2 f}{\partial x^2} \dot{x}^2 + 2  \frac{\partial^2 f}{\partial x \partial y}  \dot{x}  \dot{y} +   \frac{\partial^2 f}{\partial y^2} \dot{y}^2\rb}{
 1 + \frac{\partial f}{\partial x}^2 + \frac{\partial f}{\partial y}^2}.\label{dyn2ap}
\end{align}
These equations simplify by noticing firstly that there is no acceleration in the normal direction  to the surface
\begin{align}
\frac{1}{\sqrt{1+\frac{\partial f}{\partial x}^2+\frac{\partial f}{\partial y}^2}}
   \begin{pmatrix}
          -\frac{\partial f}{\partial x} \\
          -\frac{\partial f}{\partial y} \\
         1
       \end{pmatrix}
       \cdot
   \begin{pmatrix}
          \ddot{x} \\
          \ddot{y} \\
         \ddot{z}
       \end{pmatrix}
       =0,
 \end{align}\label{noacc}
and secondly that the vertical acceleration is given by
\beq
\ddot{z}=\ddot{x}\frac{\partial f}{\partial x}+\ddot{y}\frac{\partial f}{\partial y}+\frac{\partial^2 f}{\partial x^2}\dot{x}^2+2\frac{\partial^2 f}{\partial x\partial y}\dot{x}\dot{y}+\frac{\partial^2 f}{\partial y^2}\dot{y}^2.
\eeq
Hence, (\ref{dyn1ap})-(\ref{dyn2ap}) reduce to
\begin{align}
m\ddot{x}&=-\frac{\frac{\partial f}{\partial x} mg }{
 1 + \frac{\partial f}{\partial x}^2 + \frac{\partial f}{\partial y}^2},\\
 m\ddot{y}&=-\frac{\frac{\partial f}{\partial y} mg}{
 1 + \frac{\partial f}{\partial x}^2 + \frac{\partial f}{\partial y}^2}.
\end{align}
Now consider that the particle is pushed parallel and perpendicular to the direction of motion by forces $\boldsymbol{F}_\parallel$ and $\boldsymbol{F}_\perp$, respectively. The unit vector in the direction of motion is given by
\begin{align}
   \hat{\dot{\boldsymbol{x}}}&=\frac{1}{v}\begin{pmatrix}
          \dot{x} \\
          \dot{y} \\
         \dot{x}\frac{\partial f}{\partial x}+\dot{y}\frac{\partial f}{\partial y}
         \end{pmatrix},
  \end{align}
and the unit vector perpendicular to the direction of motion, which also lies in the tangent plane of the surface, is given by
\begin{align}
    \hat{{\boldsymbol{p}}}&=\frac{1}{v\sqrt{1+\frac{\partial f}{\partial x}^2+\frac{\partial f}{\partial y}^2}}\begin{pmatrix}
           - \frac{\partial f}{\partial x}\frac{\partial f}{\partial y}\dot{x}-\dot{y}\lb1+\frac{\partial f}{\partial y}^2\rb \\
           \frac{\partial f}{\partial x}\frac{\partial f}{\partial y}\dot{y}+\dot{x}\lb1+\frac{\partial f}{\partial x}^2\rb \\
          -\frac{\partial f}{\partial x}\dot{y}+\frac{\partial f}{\partial y}\dot{x}
         \end{pmatrix}.
  \end{align}
We take the parallel and perpendicular forces as $\boldsymbol{F}_\parallel=F_\parallel\hat{\dot{\boldsymbol{x}}}$ and $\boldsymbol{F}_\perp=F_\perp\hat{{\boldsymbol{p}}}$, without loss of generality. Therefore, the full system of dynamical equations becomes
\begin{align}
m\ddot{x}&=-\frac{\frac{\partial f}{\partial x} mg }{
 1 + \frac{\partial f}{\partial x}^2 + \frac{\partial f}{\partial y}^2}+\frac{\dot{x}F_\parallel}{v}-\frac{  \lb\frac{\partial f}{\partial x}\frac{\partial f}{\partial y}\dot{x}+\dot{y}\lb1+\frac{\partial f}{\partial y}^2\rb\rb F_\perp}{
 v\sqrt{1 + \frac{\partial f}{\partial x}^2 + \frac{\partial f}{\partial y}^2}},\label{dyn12}\\
 m\ddot{y}&=-\frac{\frac{\partial f}{\partial y} mg}{
 1 + \frac{\partial f}{\partial x}^2 + \frac{\partial f}{\partial y}^2}+\frac{\dot{y}F_\parallel}{v}+\frac{  \lb \frac{\partial f}{\partial x}\frac{\partial f}{\partial y}\dot{y}+\dot{x}\lb1+\frac{\partial f}{\partial x}^2\rb \rb F_\perp}{
 v\sqrt{1 + \frac{\partial f}{\partial x}^2 + \frac{\partial f}{\partial y}^2}}.\label{dyn22}
\end{align}
Note that rate of change of energy is given by
{\small
\beq
\frac{\mathrm{d}E}{\mathrm{d}t}=m\lb\ddot{x}\dot{x}\lb 1+\frac{\partial f}{\partial x}^2 \rb+\ddot{y}\dot{y}\lb 1+\frac{\partial f}{\partial y}^2 \rb + \frac{\partial f}{\partial x}\frac{\partial f}{\partial y}\lb\ddot{y}\dot{x}+\ddot{x}\dot{y}\rb \rb + mg \lb \frac{\partial f}{\partial x}\dot{x}+\frac{\partial f}{\partial y}\dot{y} \rb. \label{Echange}
\eeq
}
By inserting (\ref{dyn12})-(\ref{dyn22}) into (\ref{Echange}), we find
\beq
\frac{\mathrm{d}E}{\mathrm{d}t}=vF_\parallel.
\eeq
Hence, the contributions from the gravitational and perpendicular forces are zero (as expected), whereas the contribution from the parallel force is equivalent to the applied power.

For the planar Brachistochrone we have $\partial f/\partial y=\tan\alpha$ and $\partial f/\partial x=0$. In this case, the governing equations reduce to
\begin{align}
 m\ddot{x}&=\frac{\dot{x}F_\parallel}{v}-\frac{ \dot{y} F_\perp}{
 v\cos\alpha},\label{dynred1}\\
m\ddot{y}&=-mg\sin\alpha\cos\alpha+\frac{\dot{y}F_\parallel}{v}+\frac{  \dot{x}\cos\alpha F_\perp}{
 v}.\label{dynred2}
\end{align}
The corresponding energy equation is
\beq
\frac{\mathrm{d}}{\mathrm{d}t}\lb\frac{1}{2}m\lb \dot{x}^2 + \dot{y}^2\sec^2\alpha\rb + mgy\tan\alpha \rb=vF_\parallel,
\eeq
or written in terms of the rotated variable $Y=y/\cos\alpha$, we get
\beq
\frac{\mathrm{d}}{\mathrm{d}t}\lb\frac{1}{2}mv^2 + mgY\sin\alpha \rb=vF_\parallel.
\eeq

\subsection{Cylindrical polar coordinates}

Following the same steps as before, the Lagrangian in cylindrical polar coordinates is
\beq
\mathcal{L}=\frac{1}{2}m\lb \dot{r}^2\lb 1+\frac{\partial f}{\partial r}^2\rb +\dot{\theta}^2\lb r^2+\frac{\partial f}{\partial \theta}^2\rb+2\frac{\partial f}{\partial r}\frac{\partial f}{\partial \theta}\dot{r}\dot{\theta} \rb - m g f.
\eeq
The resulting Euler-Lagrange equations, after simplification, are 
\begin{align}
m \lb \ddot{r}-r \dot{\theta}^2  \rb&=-\frac{ m g \frac{\partial f}{\partial r} }{1 + \frac{\partial f}{\partial r}^2 + \frac{1}{r^2}\frac{\partial f}{\partial \theta}^2 },\\
 m \lb r\ddot{\theta} +2 \dot{r}\dot{\theta} \rb&=-\frac{ m g\frac{1}{r} \frac{\partial f}{\partial \theta}}{ 1 + \frac{\partial f}{\partial r}^2 + \frac{1}{r^2}\frac{\partial f}{\partial \theta}^2 }.
\end{align}
The unit vector in the direction of motion is given by
\begin{align}
    \hat{\dot{\boldsymbol{x}}} &= \frac{1}{v}\begin{pmatrix}
           \dot{r} \\
           r\dot{\theta} \\
          \dot{r}\frac{\partial f}{\partial r}+\dot{\theta}\frac{\partial f}{\partial \theta}
         \end{pmatrix},
  \end{align}
and the unit vector perpendicular to the direction of motion is given by
\begin{align}
    \hat{{\boldsymbol{p}}} &= \frac{1}{v\sqrt{1+\frac{\partial f}{\partial r}^2+\frac{1}{r^2}\frac{\partial f}{\partial \theta}^2}}\begin{pmatrix}
            \frac{\partial f}{\partial r} \frac{1}{r}\frac{\partial f}{\partial \theta} \dot{r} + \lb 1 +\frac{1}{r^2}\frac{\partial f}{\partial \theta}^2 \rb r\dot{\theta}\\
           - \frac{\partial f}{\partial r}\frac{1}{r} \frac{\partial f}{\partial \theta} r\dot{\theta} -\lb1+ \frac{\partial f}{\partial r}^2\rb  \dot{r} \\
         -\frac{1}{r}\frac{\partial f}{\partial \theta} \dot{r} + \frac{\partial f}{\partial r} r \dot{\theta} 
         \end{pmatrix}.
  \end{align}
Hence, the governing equations are
{\small\begin{align}
m \lb \ddot{r}-r \dot{\theta}^2  \rb&=-\frac{\frac{\partial f}{\partial r} mg }{
 1 + \frac{\partial f}{\partial r}^2 + \frac{1}{r^2}\frac{\partial f}{\partial \theta}^2}+\frac{\dot{r}F_\parallel}{v}+\frac{  \lb \frac{\partial f}{\partial r} \frac{1}{r}\frac{\partial f}{\partial \theta} \dot{r} + r\dot{\theta} \lb1+\frac{1}{r^2}\frac{\partial f}{\partial \theta}^2 \rb \rb F_\perp}{
 v\sqrt{1 + \frac{\partial f}{\partial r}^2 + \frac{1}{r^2}\frac{\partial f}{\partial \theta}^2}},\label{dyncone1}\\
 m \lb r\ddot{\theta} +2 \dot{r}\dot{\theta} \rb&=-\frac{\frac{1}{r}\frac{\partial f}{\partial \theta} mg}{
 1 + \frac{\partial f}{\partial r}^2 + \frac{1}{r^2}\frac{\partial f}{\partial \theta}^2}+\frac{r\dot{\theta}F_\parallel}{v}-\frac{  \lb  \frac{\partial f}{\partial r} \frac{1}{r}\frac{\partial f}{\partial \theta} r\dot{\theta}+\dot{r}\lb 1 + \frac{\partial f}{\partial r}^2\rb \rb F_\perp}{
 v\sqrt{1 + \frac{\partial f}{\partial r}^2 + \frac{1}{r^2}\frac{\partial f}{\partial \theta}^2}}.\label{dyncone2}
\end{align}
The rate of change of energy is given by
\beq
\begin{split}
\frac{\mathrm{d}E}{\mathrm{d}t}=&m\lb  \lb \ddot{r}-r \dot{\theta}^2  \rb \dot{r}\lb 1+ \frac{\partial f}{\partial r}^2 \rb +  \lb r\ddot{\theta} +2 \dot{r}\dot{\theta} \rb r\dot{\theta}\lb 1+ \frac{1}{r^2}\frac{\partial f}{\partial \theta}^2 \rb  \right. \\
& \left. +\frac{\partial f}{\partial r}\frac{\partial f}{\partial \theta}\lb  \lb \ddot{r}-r \dot{\theta}^2  \rb\dot{\theta} +  \lb r\ddot{\theta} +2 \dot{r}\dot{\theta} \rb \frac{\dot{r}}{r}  \rb \rb + mg \lb \frac{\partial f}{\partial r}\dot{r}+\frac{\partial f}{\partial \theta}\dot{\theta} \rb.
\end{split}\label{Echangecone}
\eeq
Again, by inserting (\ref{dyncone1})-(\ref{dyncone2}) into (\ref{Echangecone}) we get
\beq
\frac{\mathrm{d}E}{\mathrm{d}t}=v F_\parallel .
\eeq
In the case of a cone, where $\partial f/\partial\theta=0$ and $\partial f/\partial r=\tan\alpha$, (\ref{dyncone1})-(\ref{dyncone2}) reduce to
\begin{align}
m \lb \ddot{r}-r \dot{\theta}^2   \rb&=-m g \sin \alpha \cos \alpha + \frac{\dot{r}F_\parallel}{v} + F_\perp \cos \alpha \frac{r\dot{\theta}}{v} ,\label{dynredcone1}\\
m \lb r\ddot{\theta} +2 \dot{r}\dot{\theta} \rb &=  \frac{r\dot{\theta}F_\parallel}{v} - \frac{F_\perp}{\cos \alpha}\frac{\dot{r}}{v}.\label{dynredcone2}
\end{align}
The corresponding energy equation is
\beq
\frac{\mathrm{d}}{\mathrm{d}t}\lb\frac{1}{2}m\lb \dot{r}^2\sec^2\alpha + r^2\dot{\theta}^2\rb + mgr\tan\alpha \rb=v F_\parallel,
\eeq
or written in terms of the rotated variable $R=r/\cos\alpha$, we get
\beq
\frac{\mathrm{d}}{\mathrm{d}t}\lb\frac{1}{2}mv^2 + mgR\sin\alpha \rb=v F_\parallel.
\eeq

\subsection{Calculating surface gradients}

As explained in the main text, the velodrome surface is written very simply in terms of the tangent and normal coordinates, such that $z=n\tan\alpha(s)$. Therefore, to solve the dynamical equations above, we need to know how to convert the surface gradients (given in terms of Cartesian or polar coordinates) in terms of $s$ and $n$. In particular, we have the following chain rule identities
\begin{align}
\frac{\partial f}{\partial x}&=\frac{\partial f}{\partial s}\frac{\partial s}{\partial x}+\frac{\partial f}{\partial n}\frac{\partial n}{\partial x}\label{chain1},\\
\frac{\partial f}{\partial y}&=\frac{\partial f}{\partial s}\frac{\partial s}{\partial y}+\frac{\partial f}{\partial n}\frac{\partial n}{\partial y}\label{chain2},\\
\frac{\partial f}{\partial r}&=\frac{\partial f}{\partial s}\frac{\partial s}{\partial r}+\frac{\partial f}{\partial n}\frac{\partial n}{\partial r}\label{chain3},\\
\frac{\partial f}{\partial \theta}&=\frac{\partial f}{\partial s}\frac{\partial s}{\partial \theta}+\frac{\partial f}{\partial n}\frac{\partial n}{\partial \theta}.\label{chain4}
\end{align}
In the straight regions of the velodrome, these calculations are straightforward, since $s$ and $n$ are linearly related to $x$ and $y$, such that
\beq
s=\pm(y-y_i),\quad n=\pm(x-x_i),\label{sandn}
\eeq
for some constants $x_i, y_i$, $i=1,3,5$ (using the same region numbering as in the main text).

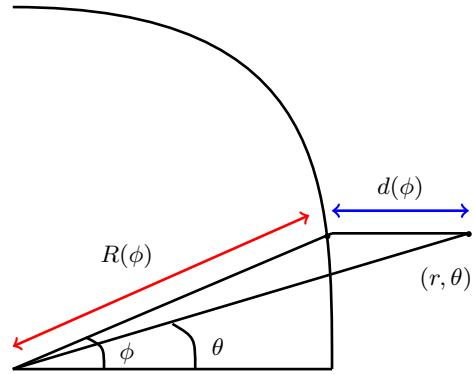
\begin{figure}
\centering
\begin{tikzpicture}[scale=0.6]
\draw [line width=1, black]  (0,8) .. controls (7,8) and (7,4) .. (7,0);
\draw [line width=1, black]  (0,0) -- (10,3);
\draw [line width=1, black]  (0,0) -- (7,3);
\draw [line width=1, red,<->]  (0,0.5) -- (6.5,3.4);
\draw [line width=1, blue,<->]  (7,3.5) -- (10,3.5);
\draw [line width=1, black]  (0,0) -- (7,0);
\draw [line width=1, black]  (7,3) -- (10,3);
\draw [line width=1, black]  (2,0) .. controls (2,0.5) .. (1.6,0.7);
\draw [line width=1, black]  (4,0) .. controls (4,0.7) .. (3.5,1);
\node at (2.5,0.35) {$\phi$};
\node at (4.5,0.5) {$\theta$};
\node at (2.5,2.5) {$R(\phi)$};
\node at (8.5,4) {$d(\phi)$};
\node at (10,2.95){\huge $\cdot$} ;
\node at (9.5,2) {$(r,\theta)$};
\node at (6.9,2.9){\huge $\cdot$} ;
\end{tikzpicture}
\caption{ Schematic diagram showing how we calculate the distance $d(\phi)$ between a point in the plane ($r$,$\theta$) and a point on the ellipse ($R(\phi)$,$\phi$). \label{aerial}}
\end{figure}

In the case of the curved sections, a little more thought is needed.  
First we note that if the curved part of the velodrome were circular, rather than elliptical, $s$ and $n$ would be linearly related to $r$ and $\theta$. Hence, as a simple approach, we consider a weakly elliptical shape (i.e. a perturbation from a circle). For an ellipse with semi-major and semi-minor axes $a(1+\epsilon)$ and $a$, respectively, the curve is given by
\beq
R(\theta)=\frac{a(1+\epsilon)}{\sqrt{(1+\epsilon)^2\cos^2 \theta + \sin^2\theta }},
\eeq
where $\epsilon\ll1$ is a small parameter. Expanding in powers of $\epsilon$, and keeping only first order terms, we get
\beq
R(\theta)\approx a(1+\epsilon \sin^2 \theta).
\eeq
Now, consider the distance from a point with polar coordinates ($r$,$\theta$), to a point on the ellipse with coordinates ($R(\phi)$,$\phi$), for some angle $\phi$, as illustrated in Fig. \ref{aerial}.  The distance between these two points is
\beq
d(\phi)=\left| \mathbf{r}-\mathbf{R} \right|,
\eeq
or equivalently,
\beq
d(\phi)=\sqrt{(r\cos \theta - R(\phi)\cos \phi)^2+(r\sin \theta - R(\phi)\sin \phi)^2}.\label{deqn}
\eeq
Considering that for a circle we have $\phi=\theta$, here we write $\phi=\theta+\epsilon \hat{\phi}$. By expanding (\ref{deqn}) in powers of $\epsilon$, we find that $d$ is given by
\beq
\begin{split}
d\approx \,& (r-a)-(a\sin^2\theta) \epsilon+\\
&\frac{a}{16(r-a)}  ( 8 r \hat{\phi}^2 +  ( r-a)(3-3\cos 4 \theta - 16\hat{\phi} \sin 2 \theta)   ) \epsilon^2.
\end{split}
\eeq
The normal coordinate $n(r,\theta)$ is given by the minimum possible value of $d$, which corresponds to some value $\hat{\phi}^*$. Clearly, it is not necessary to evaluate $\hat{\phi}^*$ to calculate $n$ to first order, since we have
\beq
n\approx (r - a) -(a   \sin^2 \theta)\epsilon.\label{norm}
\eeq
However, we calculate $\hat{\phi}^*$ here since it will be necessary when evaluating $s$ later. To find $\hat{\phi}^*$ we solve the equation
\beq
\frac{\mathrm{d} d}{\mathrm{d}\hat{\phi}}(\hat{\phi}^*)=0,
\eeq
which has solution 
\beq
\hat{\phi}^*=\frac{(r-a)}{r}\sin 2\theta.
\eeq
At a given angle $\phi$, the arclength $\mathcal{A}$ along the ellipse is given by
\begin{align}
\mathcal{A}(\phi)&=\int_0^\phi \sqrt{R(\phi)^2+R'(\phi)^2}\,\mathrm{d}\phi,\\
&\approx a\lb \phi +\epsilon\lb \frac{\phi}{2}-\frac{1}{4}\sin 2\phi \rb \rb,
\end{align}
and the tangential distance corresponds to the arclength where $\phi=\theta+\epsilon \hat{\phi}$, such that
\beq
s\approx a\lb \theta + \epsilon \lb  \frac{\theta}{2} + \frac{(3r-4a)}{4r}\sin 2\theta \rb \rb.\label{tang}
\eeq
Hence, using (\ref{norm}) and (\ref{tang}), the surface gradients in the elliptical regions (\ref{chain3})-(\ref{chain4}) can be calculated directly.

\section{Numerical optimisation\label{appB}}

In this section we briefly describe our solution method for numerical optimisation. As described in the main text, for the case of non-zero forcing in the direction parallel to the cyclist motion, no analytical solution for the optimal trajectory is available. However, a numerical solution is found by formulating a time-minimisation problem with dynamical system constraints, and following the interior point method \cite{wachter2006implementation,nocedal2006numerical}. 

In the full velodrome case, the dynamical equations are given by (\ref{dyn12})-(\ref{dyn22}) in the case of Cartesian coordinates and (\ref{dyncone1})-(\ref{dyncone2}) in the case of polar coordinates. In the simplified cases of motion on a plane or a cone, the dynamical equations are (\ref{dynred1})-(\ref{dynred2}) and (\ref{dynredcone1})-(\ref{dynredcone2}). To formulate the optimisation problem, we first discretise the state variables in time, which are either $x(t)$ and $y(t)$ in the Cartesian regions or $r(t)$ and $\theta(t)$ in the polar regions, as well as the control variable $F_\perp(t)$. Note that $F_p(t)$ is neither a state variable nor a control variable, since it is given by (\ref{hill}) in the main text. We denote the discretised state variables by $X_i$ and the control variable by $F_i$. We use a uniform discretisation in time $\delta t$ with $N$ points, such that the total time is given by $T=N\delta t$. The governing differential equations are discretised using a second order forward Euler scheme, producing a system of algebraic equations
\beq
f_j(\boldsymbol{X},\boldsymbol{F})=0,\quad j=1\ldots 2 N.\label{algeb}
\eeq
There are $2N$ equations in total since there are always two state variables (either $x$ and $y$ or $r$ and $\theta$). It should be noted that for $j=1,2$, the corresponding equations (\ref{algeb}) are precisely the initial conditions
\begin{align}
X_1&=X_0,\label{icx1}\\
\frac{1}{\delta t}\lb -\frac{3}{2}X_1+2X_2-\frac{1}{2}X_3\rb&=U_0,\label{icx2}
\end{align}
where $X_0$ and $U_0$ represent the initial position and velocity of the cyclist. 

In the case of the velodrome, as discussed in the main text, the cyclist enters between Cartesian and polar regions. Therefore, it is convenient to split (\ref{algeb}) into several partitions corresponding to the dynamical equations in each of these regions. To impose continuity between regions, the initial position and velocity (\ref{icx1})-(\ref{icx2}) at the beginning of each region must be imposed to be the same as the position and velocity at the end of the previous one. Hence, instead of solving one optimisation problem, we solve several coupled together via their initial conditions.

Another consideration that must be taken into account is that the cyclist has to remain within the bounds of the velodrome at all times. The latter constraint is written in terms of the normal coordinate $n$, which is given by (\ref{sandn}) in the case of Cartesian coordinates, and (\ref{norm}) in the case of polar coordinates, such that 
\beq
0\leq n(X_i) \leq W\cos\alpha(s(X_i)),\quad i=1\ldots N.\label{ineqcon}
\eeq

To enforce the equality constraints (\ref{algeb}) we use the quadratic penalty method. This involves placing the equality constraints as a penalty in the objective function to be maximised, which is written as
\beq
\underset{\boldsymbol{F}\in \mathbb{R}^{N}}{\mathrm{Maximise}}\quad J\lb \boldsymbol{F} \rb:=T- \mu\sum_{j=1:2N} f_j\lb\boldsymbol{X},\boldsymbol{F}\rb^2. \label{lsmin}
\eeq
The penalty parameter $\mu$ is chosen to be sufficiently large that (\ref{algeb}) is imposed accurately, but not too large that the problem becomes ill-conditioned. More details on the choice of $\mu$ are discussed by \cite{nocedal2006numerical}.
To enforce the inequality constraints (\ref{ineqcon}), we use the interior point method. This involves using logarithmic barrier functions, similar to the quadratic penalty in (\ref{lsmin}).
We make use of the IpOpt implementation of the interior point method, as discussed by \cite{wachter2006implementation}.

The optimisation problem is solved numerically using Newton's method \cite{nocedal2006numerical}, where gradients are calculated using automatic differentiation in the JuMP package \cite{dunning2017jump} of the Julia programming language \cite{bezanson2017julia}. For $N=200$, computation time on a laptop computer is around $10\un{s}$, proving a very fast method.

\section{Optimum trajectories for different starting positions\label{startps}}

\begin{figure}
\centering
\begin{tikzpicture}[scale=0.5]
\node at (0,1) {\includegraphics[width=1\textwidth]{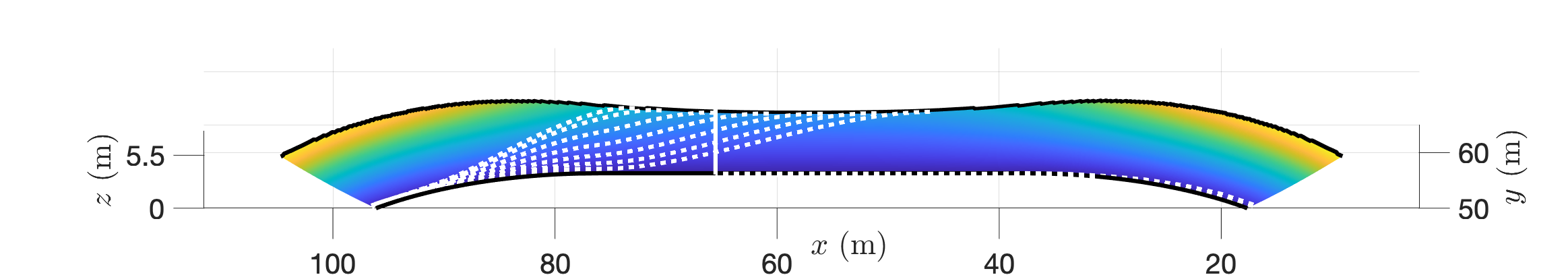}};
\draw[line width=1, <->] (-2.5,2) -- (-1.2,2);
\node at (-2,2.5) {$s(0)$};
\end{tikzpicture}
\caption{Optimum descent trajectories for different starting positions between $s(0)=7\un{m}$ and $s(0)=-20\un{m}$.
\label{delta}}
\end{figure}

\bibliographystyle{unsrt}
\bibliography{bibfile}

\end{document}